\documentclass[11pt,fleqn,twoside]{article}
\usepackage{amsfonts,amssymb,latexsym}
\makeatletter
\newcommand{\prava}{\footnotesize\it
\begin{flushright}
\begin{minipage}{18cm}
Copyright \copyright 1998 by A.M. Hamza
\end{minipage}
\end{flushright}}

\newcommand{\name}[1]{\begin{flushleft}
                       \LARGE \bf #1
                       \end{flushleft}\vspace{-3mm}}

\newcommand{\Author}[1]{\begin{flushleft}
                       \it #1 \end{flushleft}}

\newcommand{\Adress}[1]{\begin{flushleft}
                       \it #1 \end{flushleft}}

\newcommand{\Date}[1]{\begin{flushleft}
                      \small  \it #1 \end{flushleft}}

\newcommand{\ehkol}{Author \ name}
\newcommand{\ohkol}{Article \ name}
\renewcommand{\@evenhead}{
\hspace*{-3pt}\raisebox{-15pt}[\headheight][0pt]{\vbox{\hbox to \textwidth
{\thepage \hfil \ehkol}\vskip4pt \hrule}}}
\renewcommand{\@oddhead}{
\hspace*{-3pt}\raisebox{-15pt}[\headheight][0pt]{\vbox{\hbox to \textwidth
{\ohkol \hfil \thepage}\vskip4pt\hrule}}}
\renewcommand{\@evenfoot}{}
\renewcommand{\@oddfoot}{}

\newcommand{\be}{\begin{equation}}
\newcommand{\ee}{\end{equation}}
\newcommand{\ba}{\hspace*{-5pt}\begin{array}}
\newcommand{\ea}{\end{array}}

\newcommand{\ds}{\displaystyle}
\makeatother

\begin{document}

\def\theequation{\arabic{section}.\arabic{equation}}
\setcounter{page}{438}
\thispagestyle{empty}

\renewcommand{\ehkol}{A.M. Hamza}
\renewcommand{\ohkol}{Resonance Broadening Theory of Farley-Buneman
Turbulence}

\begin{flushleft}
\footnotesize \sf
Journal of Nonlinear Mathematical Physics \qquad 1998, V.5, N~4,
\pageref{hamza-fp}--\pageref{hamza-lp}.
\hfill {\sc Article}
\end{flushleft}

\vspace{-5mm}

\renewcommand{\footnoterule}{}
{\renewcommand{\thefootnote}{} \footnote{\prava}

\name{Resonance Broadening Theory of Farley-Buneman
Turbulence in the Auroral E-Region}\label{hamza-fp}

\Author{A.M. HAMZA}

\Adress{Physics Department, Center for Space Research,
University of New Brunswick, \\
P.O. Box 400, Fredericton, NB E3A 5A3, Canada}

\Date{Received June 2, 1998; Accepted August 27, 1998}

\begin{abstract}
\noindent
The conventional theory of resonance broadening for a two-species
plasma in a magnetic f\/ield is revised, and applied to an ionospheric
turbulence case. The assumptions made in the conventional theory of
resonance broadening have, in the past, led to replacing the
frequency $\omega$ by $\omega+ik_{\perp}^{2}D^{*}$ in the resonant
part of the linear dielectric function to obtain the nonlinear
dielectric function. Where $D^{*}$ is an anomalous dif\/fusion
coef\/f\/icient due solely to wave scattering of the particle orbits. We
show that in general these assumptions are not valid, and
consequently the straightforward substitution of frequencies is not 
legitimate. We remedy these problems and derive expressions for the
time-dependent components of the dif\/fusion tensor. The improved resonance
broadening theory is developed in the context of an ionospheric problem,
namely that of the Farley-Buneman turbulence in the auroral E-region. A
kinetic description of the electrons is used. A general expression for the
nonlinear dielectric function is derived in the special case where no
parallel electric f\/ield is present, and the dif\/ferences with the conventional
dispersion relation are discussed.
\end{abstract}

\section{Introduction}
\label{sec-introduction}

{\advance\baselineskip-0.25pt

The problem of wave-particle interaction has always played a critical
role when one tries to understand the saturation mechanisms for plasma
instabilities. Linear theory, which is a single wave theory, which does not
take into account the wave-particle interaction fails to conserve energy and
momentum, and consequently does not and should not predict
sa\-tu\-ra\-tion. This is 
unphysical. But it is important to remember that the linear theory ceases to
be valid after a trapping time; the time neccessary to a charged particle to
bo\-un\-ce near the bottom of a potential well of a f\/inite-amplitude wave.
Consequently, the pre\-dictions of linear theory beyond the trapping time are
not valid. A f\/irst remedy to this fun\-damental problem is provided by the so
called Quasilinear theory, a weak turbulence theory which takes into
consideration the wave-particle interaction and requires a slow time
dependence of the background particle distribution function. The ensemble
averaged distribution function is then a solution to a Fokker-Planck equation.
In the absence of sources and sinks, quasilinear theory is then described
through a  dif\/fusion equation which predicts saturation of the instabilities
when the background distribution function becomes constant, or ``plateaus'',
along the dif\/fusion paths which are the characteristics~of~the par\-tial
dif\/ferential equation. However, in the presence of sources and
sinks, such as a back\-ground electric f\/ield or collisions, there is no
saturation, since the sources and sinks tend~to destroy the plateau, and
consequently the unstable waves keep on growing until they be\-co\-me large
enough for the other nonlinear processes to enter the picture. This~is the
case of the problem we have elected to address, namely that of the modif\/ied
two stream Farley-Buneman instability, where the source of free energy is the
electrojet and the sink~is due to the collisions of the electrons with the
background neutrals. In other words quasilinear theory does not provide an
ultimate saturation mechanism much needed to describe steady state turbulence
and predict reasonable saturation amplitudes. Nevertheless the quasi\-linear
theory has stood the test of time when it comes to predicting the onset of
in\-sta\-bilities to a certain degree of accuracy. A f\/irst attempt to improve on
the quasilinear theory was described by Dupree~[5] and Weinstock~[20]
for an unmagnetized plasma and~by Dupree~[6] and 
 Dum and Dupree~[7] for a magnetized plasma.  Dupree~[5]
derived a perturbation theory based on the knowledge of the electric f\/ield
which allowed him to f\/ind the exact particle orbits which were then used in
the perturbation solutions of the Vlasov equation. The principal result of
this improved perturbation theory is a broadening of the wave-particle
resonance function, which in the conventional quasilinear theory is
$\delta(\omega-{\bf k}\cdot{\bf v})$. This result has been the subject of
many debates on strong turbulence theories in the past two decades. The
validity of some of the assumptions made by Dupree~[5] was questioned
by a number of authors (Cook and Sanderson~[3], Rolland~[15]). A
number of authors (Salat~[18], Ishihara and Hirose~[10], 
Ishihara {\it et al.}~[11, 12] have since
taken the task to study the problem of resonance broadening very thoroughly,
and were able to show successfully the shortcomings of the conventional
resonance broadening approach. They all addressed the resonance broadening
problem in the case of unmagnetized plasmas, as well as the case related to
drift waves in a shear magnetic f\/ield because of the mathematical
tractability and its reduction to a form similar to the unmagnetized case.
Kleva and Drake~[21] addressed the problem of stochastic ${\bf
E}\times{\bf B}$ particle transport using a dynamical system approach. More
recently  Kleva~[14] investigated the problem of energy transport in a
magnetically conf\/ined plasma by ${\bf E}\times{\bf B}$ f\/low generated by a
spectrum of electrostatic waves stil using a dynamical system's approach. He
found like  Ishihara {\it et al.}~[12] that the dif\/fusion coef\/f\/icient scales
like $E^{4/3}$ instead of $E^{2}$ as predicted by the conventional quasilinear
theory. The problem addressed by Dupree~[6] and Dum and Dupree~[7],
that of the broadening of the wave-particle resonance in the presence 
of background magnetic f\/ield, has not been addressed since probably
because of its complexity.

In this paper we have investigated the problem of resonance broadening for a
turbulent magnetized plasma and compared our results to the classical
calculation of Dum and Dupree~[7]. We have investigated the
implications of the improved results on the Farley-Buneman instability that
occurs in ionospheric plasmas, and shown that one cannot obtain the nonlinear
dielectric function by just substituting the frequency $\omega$ by $\omega
+ik_{\perp}^{2}D^{*}$ in the resonant part of the dielectric function. In
Section~2 we describe the mathematical model used to
investigate the Farley-Buneman instability, and derive the dif\/ferent
components of the dif\/fusion tensor. In Section~3 we
derived the generalized dispersion relation, and f\/inally in Section~4
we  properly reinterpret the thresholds
conditions for the Farley-Buneman instability in the absence of a parallel
electric f\/ield.

}

\section{The Mathematical Model}\label{sec-model}

\subsection{The Quasi-Linear Approximation}
\label{sec-quasilinear}

\subsubsection{The Ion Description}\label{sec-ion}

The ions are assumed to be highly collisional in the region of
interest, namely the auroral E-region, and unmagnetized. The ion
convection is also assumed to be negligible, i.e., the nonlinear ion
terms are neglected. 

With these assumptions a linear f\/luid model is adopted for the ions,
and is best described by the linearized momentum and continuity equations.
Following Sudan~[19] we write 
\[
m_{i}{\partial{\bf\delta\bf v}_{i}\over\partial
t}=-e{\bf\nabla}\phi-\frac{T_{i}}{n_{0}}{\bf\nabla}\delta
n-\nu_{in}m_{i}{\bf\delta\bf v}_{i}, \label{eq:i1}
\]
\begin{equation}
{\partial\delta n\over\partial t}+n_{0}{\bf\nabla}\cdot{\bf\delta\bf v}_{i}=0
\label{eq:i2}
\end{equation}
operating on the continuity equation (\ref{eq:i2}) with
$\ds \left({\partial\over\partial t}+\nu_{in}\right)$ leads to
\be
\left({\partial\over\partial t}+\nu_{in}\right)
{\partial\delta n\over\partial t}  =  -n_{0}{\bf\nabla}\cdot\left(
{\partial\over\partial t}+\nu_{in}\right){\bf\delta\bf v}_{i} 
                                   =  n_{0}{\bf\nabla}\cdot\left\{
\frac{e}{m_{i}}{\bf\nabla}\phi+\frac{T_{i}}{n_{0}m_{i}}{\bf\nabla}\delta n
\right\}.   \label{eq:i3}
\ee
Equation (\ref{eq:i3}) allows us to express the density f\/luctuation in terms
of the electric f\/ield. Taking the Fourier transform in space and time of
equation (\ref{eq:i3}) leads to
\begin{equation}
\left(\omega^{2}-k^{2}\frac{T_{i}}{m_{i}}+i\nu_{in}\omega\right)\delta n_{\bf
k\omega}=\frac{n_{0}e}{m_{i}}k^{2}\phi_{\bf k\omega}.  \label{eq:i4}
\end{equation}
This equation will eventually be used in Poisson's equation to determine the
dispersion relation.

\subsubsection{The Electron Description}\label{sec-electron}

The electrons are described by the Vlasov equation with a relaxation
model for collision operator. The electron distribution function satisf\/ies the
following equation 
\be
\ba{l}
\ds \left({\partial \over \partial t}+{\bf
v}\cdot{\bf\nabla}-\frac{e}{m_{e}}\left({\bf E}+\frac{\bf
v}{c}\times{\bf B}\right)\cdot{\bf \nabla}_{\bf v}\right)f_{e}
({\bf x},{\bf v},t)=
\vspace{3mm}\\
\ds \qquad  -\nu_{en}\left(f_{e}({\bf x},{\bf v},t)-\frac{n({\bf
x},t)}{n_{0}}f_{0}\right),
\ea
\label{eq:e1}
\ee
where 
\be
\ba{l}
\ds
{\bf E}({\bf x},t)   =  {\bf E}_{0}+{\bf\delta\bf E}({\bf x},t),
\vspace{3mm}\\
\ds n({\bf x},t)  =  \int d{\bf v}f_{e}({\bf x},{\bf v},t).
\ea
\label{eq:e2}
\ee

The electric f\/ield has been separated into two parts;
${\bf E}_{0}$ represents the electrojet background electric f\/ield, and
${\bf\delta E}$ the f\/luctuating f\/ield.

This leads to the following distribution function, which is in
turn separated into a weakly space and tine dependent average distibution
function $\langle f_{e}({\bf x},{\bf v},t)\rangle$, and a f\/luctuating part
$\delta f_{e}({\bf x},{\bf v},t)$, that is
\begin{equation}
f_{e}({\bf x},{\bf v},t)=\langle f_{e}({\bf x},{\bf v},t)\rangle+
\delta f_{e}({\bf x},{\bf v},t).  \label{eq:e3}
\end{equation}

We should point out that both the electric f\/ield and the
distribution function decompositions are exact.
One can easily write equation (\ref{eq:e1}) in a compact form
\begin{equation} 
\left({\partial \over \partial t}+i{\cal L}({\bf x},{\bf
v},t)\right) f_{e}({\bf x},{\bf v},t)=\frac{e}{m_{e}}{\bf\delta\bf
E}({\bf x},t)\cdot\nabla_{\bf v}f_{e}({\bf x},{\bf v},t),
\label{eq:e4}
\end{equation} 
where the operator ${\cal L}$ is given by:
\[
i{\cal L}({\bf x},{\bf v},t)={\bf v}\cdot{\bf\nabla}-\frac{e}{m_{e}}\left({\bf
E}_{0}+\frac{\bf v}{c}\times{\bf B}\right)\cdot{\bf \nabla}_{\bf
v}+\nu_{en}-\nu_{en} \frac{f_{0}}{n_{0}}\int d{\bf v}.  \label{eq:e5}
\]

Note that the collision operator is a linear operator, and consequently its
incorporation into the collisionless Vlasov equation is not as complex as if
we would have used a Fokker-Plank or any more sof\/isticated collision operator. 

Equation (\ref{eq:e4}) can be solved using dif\/ferent techniques. In the
quasi-linear approximation one considers only the wave particle interaction
and neglects any other ef\/fects, such as wave coupling and radiation ef\/fects
which are nonlinear ef\/fects of higher order in a perturbation analysis based
on the amplitude of the wave electric f\/ield.

To solve the quasi-linear problem we start by ensemble averaging the collisional
Valsov equation to obtain
\be
\ba{l}
\ds \left({\partial \over \partial t}+\nu_{en}+{\bf
v}\cdot{\bf\nabla}-\frac{e}{m_{e}}\left({\bf E}_{0}+\frac{\bf v}{c}\times{\bf
B}\right)\cdot{\bf\nabla}_{\bf v}\right)\langle f_{e}\rangle
\vspace{3mm}\\
\ds \qquad =  \nu_{en}f_{0} +
\frac{e}{m_{e}}{\bf\nabla}_{\bf v}\cdot\langle {\bf\delta\bf E}\delta
f_{e}\rangle.
\ea     \label{eq:e6}
\ee

The next step consists of substituting the expression for the distribution
function into the collisional Vlasov equation to obtain
\be
\ba{l}
\ds \left({\partial \over \partial t}+\nu_{en}+{\bf
v}\cdot{\bf\nabla}-\frac{e}{m_{e}}\left({\bf E}_{0}+\frac{\bf v}{c}\times{\bf
B}\right)\cdot{\bf\nabla}_{\bf v}\right)(\langle f_{e}\rangle+\delta
f_{e}) -\frac{e}{m_{e}}{\bf\delta\bf E}\cdot{\bf\nabla}_{\bf v}\langle f_{e}
\rangle
\vspace{3mm}\\
\ds \qquad = \nu_{en}f_{0}+\nu_{en}\frac{f_{0}}{n_{0}}\delta n+\frac{e}{m_{e}}
{\bf\nabla}_{\bf v}\cdot({\bf\delta\bf E}\delta f_{e}),
\ea
\label{eq:e7}
\ee
where $\delta n$ is def\/ined as
\[
\delta n=\int d{\bf v} \delta f_{e}.   \label{eq:e8}
\]

Taking the dif\/ference between equation
(\ref{eq:e7}) and equation (\ref{eq:e6}) leads to
\be
\ba{l}
\ds
\left({\partial \over \partial t}+\nu_{en}+{\bf
v}\cdot{\bf\nabla}-\frac{e}{m_{e}}\left({\bf E}_{0}+\frac{\bf v}{c}\times{\bf
B}\right)\cdot{\bf\nabla}_{\bf
v}\right)\delta f_{e} =
\vspace{3mm}\\
\ds \qquad -\frac{e}{m_{e}}{\bf\delta\bf
E}\cdot{\bf\nabla}_{\bf v}\langle
f_{e}\rangle-\nu_{en}\frac{f_{0}}{n_{0}}\delta
n=\frac{e}{m_{e}}{\bf\nabla}_{\bf v}\cdot\left\{{\bf\delta\bf E}\delta
f_{e}-\langle {\bf\delta\bf E}\delta f_{e}\rangle\right\}.
\ea
\label{eq:e9}
\ee

The quasi-linear approximation consists of neglecting the right hand side of
equation~(\ref{eq:e9}) which describes the nonlinear mode coupling terms. When
this term is neglected, the equation to lowest order in the electric f\/ield
becomes
\[
\left({\partial \over \partial t}+\nu_{en}+{\bf
v}\cdot{\bf\nabla}-\frac{e}{m_{e}}\left({\bf E}_{0}+\frac{\bf v}{c}\times{\bf
B}\right)\cdot{\bf\nabla}_{\bf v}\right)\delta f_{e}=\nu_{en}\frac{f_{0}}{n_{0}}
\delta n+\frac{e}{m_{e}}{\bf\delta\bf E}\cdot{\bf\nabla}_{\bf v}\langle
f_{e}\rangle.  \label{eq:e10}
\]
This equation can be solved formally using Fourier transforms and def\/ining the
following operator 
\[
{\bf G}_{{\bf k}\omega}({\bf v})=\left\{-i\omega+\nu_{en}+i{\bf k}\cdot{\bf
v}-\frac{e}{m_{e}}\left({\bf E}_{0}+\frac{\bf v}{c}\times{\bf
B}\right)\cdot{\bf\nabla}_{\bf v}\right\}^{-1}. \label{eq:e11}
\]
The expression for the f\/luctuating distribution function can then be written in
the following form
\begin{equation}
\delta f_{e{\bf k}\omega}({\bf v})=\frac{e}{m_{e}}{\bf\delta\bf
E}_{{\bf k}\omega}\cdot{\bf G}_{{\bf k}\omega}{\bf\nabla}_{\bf v} \langle
f_{e}\rangle+\nu_{en}\frac{{\delta n}_{{\bf k}\omega}}{n_{0}}{\bf G}_{{\bf k}
\omega}f_{0}.
\label{eq:e12} 
\end{equation}
Integrating over velocity space one obtains an expression for the density
f\/luctuation which when substituted into expression (\ref{eq:e12}) for the
f\/luctuating distribution function leads to
\begin{equation}
\delta f_{e{\bf k}\omega}=\frac{e}{m_{e}}{\bf\delta\bf E}_{{\bf k}\omega}\cdot
{\bf G}_{{\bf k}\omega}{\bf\nabla}_{\bf v}
\langle f_{e}\rangle+\nu_{en}\frac{\frac{e}{m_{e}}{\bf\delta\bf E}_{{\bf k}
\omega}\cdot\int
d{\bf v}{\bf G}_{{\bf k}\omega}{\bf\nabla}_{\bf v}\langle f_{e}\rangle}{n_{0}-
\nu_{en}\int
d{\bf v}{\bf G}_{{\bf k}\omega}f_{0}}{\bf G}_{{\bf k}\omega}f_{0}
\label{eq:e13}
\end{equation}
when substituting expression (\ref{eq:e13}) into the equation governing
the evolution of the average distribution equation one obtains
\[
{\partial\langle f_{e}\rangle\over\partial t}={\partial\over\partial{\bf
v}}\cdot{\underline{\underline{\bf D}}}\cdot{\partial\langle
f_{e}\rangle\over\partial{\bf v}}-{\partial\over\partial{\bf
v}}\cdot{\underline{\bf F}}\langle f_{e}\rangle.  \label{eq:e14}
\]
In other words one obtains a Fokker-Planck equation with the dif\/fusion and drag
coef\/f\/i\-cients def\/ined as follows
\[
\ba{l}
\ds {\underline{\underline{\bf D}}}   =
\frac{e^{2}}{m_{e}^{2}}\left\langle{\bf\delta\bf E}{\bf\delta\bf E}{\bf
G}\right\rangle,
\vspace{3mm}\\
\ds {\underline{\bf F}}\langle f_{e}\rangle  =  -\frac{e}{m_{e}}{\bf
E}_{0}\langle f_{e}\rangle-\frac{e^{2}}{m_{e}^{2}}\nu_{en}
\left\langle{\bf\delta\bf E}{\bf\delta\bf E}{\bf G}\cdot\frac{\int d{\bf v}{\bf
G}{\partial\langle f_{e}\rangle\over\partial{\bf v}}}{n_{0}-\nu_{en}\int d{\bf
v}{\bf G}f_{0}}f_{0}\right\rangle. \label{eq:e15}
\ea
\]
It is clear that the drag term in the Fokker-Planck equation is solely due to
the background electric f\/ield of the electrojet and to the electron-neutral
collisions.

In the absence of sources or collisions the conventional quasi-linear theory
predicts a saturation of the f\/luctuating f\/ields when the distribution function
becomes constant along the dif\/fusion paths. That is, the quasi-linear theory
predicts a zero growth rate. This however, is valid only when there are no
sources or collisions; when such ef\/fects are present, like in our case, the
distribution function never plateaus along the dif\/fusion paths since the source
and the collisions tend to destroy the plateau, which in turn leads to a non
zero growth rate; the oscillation amplitude continues to grow until
nonlinear processes enter the picture. In other words, there is no ultimate
saturation of the f\/luctuations through the quasilinear process.

This leads us to conclude that the conventional quasi-linear theory is not
the ultimate stabilization mechanism, and improvements on the theory are
needed. One possible theory that has been suggested, as a f\/irst attempt to
remedy the problem from which the quasilinear theory suf\/fers, is the resonance
broadening theory which was f\/irst introduced by Dupree~[5],
and applied by a number of authors, see for example Sudan~[19],
Robinson~[16], and Robinson and Honary~[17], to the problem of
irregularities in the E region. 

In the next section we shall develop the resonance broadening in some details.
We will also discuss, to a certain extent, the validity limits of the
conventional resonance broadening. We remedy some of the problems, and
show the shortcomings of this wave particle interaction, as well as the need to
provide for a complete theory of plasma turbulence. The latter will be addressed
in a companion paper describing a fully selfconsistent kinetic theory for the
Farley-Buneman instability.

\subsection{The Resonance Broadening Approximation}
\label{sec-resonance}

In this section we shall address the problem of wave particle
interaction in a magnetized plasma through a nonlinear formalism that includes
the nonlinear ef\/fects of the waves on the particle orbits, but not vice versa.
The electric f\/ield and the corresponding distribution function are decomposed
according to equations (\ref{eq:e2}) and (\ref{eq:e3}).

The fundamental assumption and the goal of the resonance broadening theory is
to evaluate the modif\/ication of the quasilinear resonance between the
particles and waves. This theory neglects the coherent contributions
which arise from the coupling between the waves and the background
oscillations, as well as interactions between the background oscillations.

We def\/ine the following operator ${\cal L}$ is then def\/ined as follows
\[
i{\cal L}({\bf x},{\bf v},t)={\bf v}\cdot{\bf\nabla}-\frac{e}{m_{e}}\left({\bf
E}_{0}+{\bf\delta\bf E}+\frac{\bf v}{c}\times{\bf B}\right)\cdot{\bf \nabla}_{\bf
v}+{\cal O}^{coll}({\bf x},{\bf v},t),  \label{eq:r2}
\]
where the collision operator ${\cal O}^{coll}$ is given by
\[
{\cal O}^{coll}({\bf x},{\bf v},t)=\nu_{en}-\nu_{en}\frac{f_{0}(\bf
v)}{n_{0}}\int d{\bf v}' \label{eq:r2a}
\]
which allows us to rewrite equation (\ref{eq:e1}) in a compact form
\[
\left({\partial\over\partial t}+i{\cal L}({\bf x},{\bf v},t)\right)f_{e}
({\bf x},{\bf v},t)=0.  \label{eq:zr0}
\]

The equation describing the evolution of the ensemble averaged electron
distribution function is given by
\[
\left({\partial\over\partial t}+i{\cal L}_{0}({\bf x},{\bf v},t)\right)
\langle f_{e}({\bf x},{\bf v},t)\rangle=
\frac{e}{m_{e}}{\bf\nabla}_{\bf v}\cdot\langle{\bf\delta\bf E}({\bf x},t)
\delta f_{e}({\bf x},{\bf v},t)\rangle,  \label{eq:zr1}
\]
where ${\cal L}_{0}$ is a linear operator def\/ined by
\[
i{\cal L}_{0}({\bf x},{\bf v},t)={\bf
v}\cdot{\bf\nabla}+\left(-\frac{e}{m_{e}}{\bf
E}_{0}+{\bf\Omega}_{e}\times{\bf v}\right)\cdot{\bf\nabla}_{\bf v}
+\nu_{en}-\nu_{en} \frac{f_{0}({\bf v})}{n_{0}}\int d{\bf v}'.
\label{eq:zr2}
\]

The equation governing the evolution of the f\/luctuating part of the
electron distribution function can be written as follows 
\begin{equation}
\left({\partial\over\partial t}+i{\cal L}({\bf x},{\bf v},t)\right)
\delta f_{e}({\bf x},{\bf v},t)=
\frac{e}{m_{e}}{\bf\delta\bf E}({\bf x},t)\cdot{\bf\nabla}_{\bf v}\langle
f_{e}({\bf x},{\bf v},t)\rangle, \label{eq:r3}
\end{equation}
where as one can see the nonlinear term ${\bf\delta\bf E}\delta f_{e}$. This
term will take into account the ef\/fects of the waves on the electrons in
the resonance broadening approximation, but will not address or investigate
the ef\/fects of the electrons on the waves, neither does it take into
consideration the wave-wave interaction. The ``Resonance Broadening''
approximation assumes that the quadratic nonlinearity leads to a nonlinear
correction to the particle orbits, which should be taken into account when the
f\/ields become large enough.  

Equation (\ref{eq:r3}) can be solved using a
Green's function analysis, i.e., we def\/ine the Green's function $\cal G$ which
satisf\/ies 
\begin{equation} 
\left({\partial\over\partial t}+i{\cal L}({\bf x},{\bf
v},t)\right) {\cal G}({\bf x},{\bf v},t;{\bf x}',{\bf
v}',t')={\bf\delta}({\bf x}- {\bf x}'){\bf\delta}({\bf v}-{\bf
v}'){\bf\delta}(t-t'). \label{eq:r4}
\end{equation} 
Following Ishihara {\it et al.}~[12] the operator $i{\cal L}
$ can be rewritten as follows
\be
\ba{l}
\ds i{\cal L}      =  i{\cal L}_{0}+i{\cal L}_{1},
\vspace{3mm}\\
\ds i{\cal L}_{0}  =  {\bf v}\cdot{\bf\nabla}+\left(-\frac{e}{m_{e}}{\bf
E}_{0}+{\bf\Omega}_{e}\times{\bf v}\right)\cdot{\bf\nabla}_{\bf v}
+{\cal O}^{coll}, \vspace{3mm}\\
\ds i{\cal L}_{1}  =  -\frac{e}{m_{e}}{\bf\delta\bf E}({\bf x},t)
\cdot{\bf\nabla}_{\bf v}. \label{eq:xr4}
\ea
\ee
The electron distribution function is then obtained 
\[
f_{e}({\bf x},{\bf v},t)=\int d{\bf x}'d{\bf v}'{\cal G}({\bf x},
{\bf v},t;{\bf x}',{\bf v}',t_{0})f_{e}({\bf x}',{\bf v}',t_{0}).
\label{eq:xr5}
\]

Before we get into the details of the derivation, we should point out that
throughout the calculation we will omit the collisions. We will discuss the
impact of collisions on the dispersion relation when we apply the results to the
case of Farley-Byneman turbulence using the relaxation model for the collision
operator already shown above.

We now separate the Green's function into two parts; an unperturbed part 
${\cal G}^{(0)}$ and a perturbed one ${\cal G}^{(1)}$, i.e.,
\begin{equation}
{\cal G}={\cal G}^{(0)}+{\cal G}^{(1)} \label{eq:xr6}
\end{equation}
satisfying the following equations
\begin{equation} 
\left({\partial\over\partial t}+i{\cal L}_{0}({\bf x},{\bf
v},t)\right) {\cal G}^{(0)}({\bf x},{\bf v},t;{\bf x}',{\bf
v}',t')={\bf\delta}({\bf x}- {\bf x}'){\bf\delta}({\bf v}-{\bf
v}'){\bf\delta}(t-t'). \label{eq:xr7}
\end{equation}
The solution to equation (\ref{eq:xr7}) can be written in the following formal
form 
\begin{equation}
{\cal G}^{(0)}({\bf x},{\bf v},t;{\bf x}',{\bf v}',t')=\Theta(t-t')
{\bf\delta}({\bf x}(t)-{\bf x}'(t')){\bf\delta}({\bf v}(t)-{\bf
v}'(t')), \label{eq:r5}
\end{equation}
where $\Theta(t-t')$ is the Heavyside step function.
Substracting equation (\ref{eq:xr7}) from equation~(\ref{eq:r4}) leads to the
following expression for the perturbed part of the Green's function
\[
\ba{l}
\ds {\cal G}^{(1)}({\bf x},{\bf v},t;{\bf x}',{\bf
v}',t')
\vspace{3mm}\\
\ds \quad =    -i\int_{t'}^{t} dt''\int d{\bf x}''\int {\bf
v}'' {\cal G}^{(0)}({\bf x},{\bf v},t;{\bf x}'',{\bf v}'',t''){\cal
L}_{1} ({\bf x}'',{\bf v}'',t'')
    {\cal G}({\bf x}'',{\bf v}'',t'';{\bf x}', {\bf v}',t').
 \label{eq:xr7a}
\ea
\]
Def\/ine the average Green's function $G({\bf x},{\bf v},t;t')$ as follows
\[
G({\bf x},{\bf v},t;t')=\int d{\bf x}'\int d{\bf v}' \;\langle
{\cal G}({\bf x},{\bf v},t;{\bf x}',{\bf v}',t')\rangle,
\label{eq:xr8}
\]
where $\langle\cdots\rangle$ represent an ensemble average.
Taking the ensemble
average of equation (\ref{eq:r4}) and integrating over ${\bf x}'$
and ${\bf v}'$ leads to the following equation
\begin{equation}
{\partial G({\bf x},{\bf v},t;t')\over\partial t}+\int d{\bf x}'\int d{\bf
v}'\; \langle i{\cal L}({\bf x},{\bf v},t){\cal G}({\bf x},{\bf v},t;{\bf
x}',{\bf v}',t')\rangle=\delta(t-t').  \label{eq:xr9}
\end{equation}
Substituting equations (\ref{eq:xr4}), (\ref{eq:xr6}), (\ref{eq:r5}) and
(\ref{eq:xr7}) into equation (\ref{eq:xr9}) we obtain
\be
\ba{l}
\ds {\partial G({\bf x},{\bf v},t;t')\over\partial t}+\int d{\bf x}'
\int d{\bf v}'\int_{t'}^{t} dt''\int d{\bf x}''\int d{\bf v}''
\vspace{3mm} \\
\qquad \ds  \times\langle {\cal L}_{1}({\bf x},{\bf v},t)
{\cal G}^{(0)}({\bf x},{\bf v},t;{\bf x}'',{\bf v}'',t''){\cal L}_{1}
({\bf x}'',{\bf v}'',t''){\cal G}({\bf x}'',{\bf v}'',t'';{\bf x}',
{\bf v}',t')\rangle. \label{eq:xr10}
\ea
\ee
Following the work of Ishihara {\it et al.}~[12] and references therein we can
simplify equation~(\ref{eq:xr10}) after making the following approximations
in order to evaluate the second term on the left hand side. We assume that the
process under consideration is Gaussian, which leads to
\be
\ba{l}
\ds
\!\!\int d{\bf x}'\!\int d{\bf v}'\; \langle {\cal L}_{1}({\bf x},{\bf v},
t){\cal G}^{(0)}({\bf x},{\bf v},t;{\bf x}'',{\bf v}'',t''){\cal
L}_{1}({\bf x}'',{\bf v}'',t''){\cal G}({\bf x}'',{\bf v}'',t'';{\bf
x}',{\bf v}',t')\rangle
\vspace{3mm}\\
\ds \qquad =
  \langle {\cal L}_{1}({\bf x},{\bf v},
t){\cal G}^{(0)}({\bf x},{\bf v},t;{\bf x}'',{\bf v}'',t''){\cal
L}_{1}({\bf x}'',{\bf v}'',t'')\rangle G({\bf x}'',{\bf
v}'',t'';t').  \label{eq:xr11}
\ea       \hspace{-11.38pt}
\end{equation}
The direct interaction approximation makes a further approximation and replaces 
${\cal G}^{(0)}$ in the right hand side of equation (\ref{eq:xr11}) by ${\cal
G}$. Finally, the last approximation consists of replacing $G({\bf x}'',
{\bf v}'',t'';t')$ by $G({\bf x},{\bf v},t;t')$, and ${\cal
L}_{1}({\bf x}'',{\bf v}'',t'')$ by ${\cal
L}_{1}({\bf x},{\bf v},t'')$ to obtain
\be
\ba{l}
\ds
{\partial G({\bf x},{\bf v},t;t')\over\partial t}+\int_{t'}^{t}
dt'' \int d{\bf x}''\int d{\bf v}''
\vspace{3mm}\\
\ds \qquad  \times\langle {\cal L}_{1}({\bf x},{\bf v},
t){\cal G}({\bf x},{\bf v},t;{\bf x}'',{\bf v}'',t''){\cal
L}_{1}({\bf x},{\bf v},t'')\rangle G({\bf x},{\bf v},t;t')=0.
\ea
\label{eq:xr12} 
\ee
Noting that ${\cal L}_{1}({\bf x},{\bf v},t)$ is given by equation
(\ref{eq:xr4}) we can rewrite equation (\ref{eq:xr12}) as a dif\/fusion equation.
However before we do this we will assume that the ensemble averaged
distribution function changes very slowly through secular changes of
the integrals (constants) of the unperturbed motion, i.e., we assume that
\[
\langle f_{e}({\bf x},{\bf v}, t)\rangle=\langle f_{e}(\tilde{\bf x},
v_{\perp},v_{\parallel},t)\rangle,  \label{eq:xr12a}
\]
where $\tilde{\bf x}$ represent the guiding center coordinates of the
electrons and $v_{\perp}$, $v_{\parallel}$ the perpendicular and parallel
velocity components, respectively. This assumption allows us to express the
velocity gradient in terms of gradients in the guiding center
coordinates. The xpression for $i{\cal L}_{1}$ is as follows
\[
{\bf\delta\bf E}_{\bf k}(t)\cdot{\bf\nabla}_{\bf v}=-i\Phi_{\bf k}(t){\bf
k}\cdot{\bf\nabla}_{\bf v} \label{eq:xr12z}
\]
which requires
\begin{equation}
{\bf k}\cdot{\bf\nabla}_{\bf v}={\bf k}_{\perp}\cdot{\bf
v}_{\perp}\frac{1}{v_{\perp}}{\partial\over\partial
v_{\perp}}+k_{\parallel}{\partial\over\partial v_{\parallel}}+\frac{{\bf
k}\times\hat{\bf z}}{\Omega_{e}}\cdot\tilde{\bf\nabla},  \label{eq:xr12b}
\end{equation}
where $\tilde{\bf\nabla}$ represents the gradient with respect to the guiding
center coordinates $\tilde{\bf x}$. Note that the last term on the right
hand side of equation (\ref{eq:xr12b}) comes from the ${\bf\delta\bf E}
\times{\bf B}$ contribution. The dif\/fusion equation can then be written in a
compact form as follows
\[
{\partial G({\bf X},t;t')\over\partial
t}=\sum_{\alpha,\beta}{\partial\over\partial
X_{\alpha}}D_{\alpha\beta}{\partial\over\partial X_{\beta}},  \label{eq:xr13}
\]
where ${\bf X}$ represents the guiding center coordinates as well as the
perpendicular and parallel velocity components, i.e.,
\[
{\bf X}\equiv\left\{\tilde{\bf x},v_{\perp},v_{\parallel}\right\}
\label{eq:xr13a}
\]
and where by def\/inition the dif\/fusion coef\/f\/icients are given by
\[
D_{\alpha\beta}=\frac{1}{2}\frac{d}{dt}\langle \Delta X_{\alpha}\Delta
X_{\beta}\rangle  \label{eq:xr13b}
\]
more explicitly the dif\/fusion coef\/f\/icients are given by (when setting
$t'=t_{0}=0$)
\[
\ba{l}
\ds \tilde{\bf D}_{\perp\perp}   =  \frac{1}{2}\frac{d}{dt}\langle
{\bf\delta\bf\tilde{x}}_{\perp}(t){\bf\delta\bf\tilde{x}}_{\perp}(t)\rangle
\vspace{3mm} \\
\ds \phantom{
\tilde{\bf D}_{\perp\perp}} =
\int_{0}^{t} dt''\int d{\bf x}''\int d{\bf v}'' \;\langle {\bf v}_{E}({\bf
x},t){\cal G}({\bf x},{\bf v},t;{\bf x}'',{\bf v}'',t''){\bf v}_{E}({\bf
x},t'')\rangle,
\ea
\]
\[
\ba{l}
\ds {\bf D}_{\perp v_{\parallel}}      =  \frac{1}{2}\frac{d}{dt}\langle
{\bf\delta\bf\tilde{x}}_{\perp}(t)\delta v_{\parallel}(t)\rangle
\vspace{3mm}\\
\ds  \phantom{{\bf D}_{\perp v_{\parallel}}} =  -\frac{e}{m_{e}}\int_{0}^{t}
dt''\int  d{\bf x}''\int d{\bf v}'' \; \langle {\bf v}_{E}({\bf x},t){\cal
G}({\bf x}, {\bf v},t;{\bf x}'',{\bf v}'',t''){\bf\delta\bf
E}_{\parallel}({\bf x},t'')\rangle,
\ea
\]
\[
\ba{l}
\ds {\bf D}_{v_{\parallel}v_{\parallel}}   =
 \frac{1}{2}\frac{d}{dt}\langle
\delta v_{\parallel}(t)\delta v_{\parallel}(t)
\vspace{3mm} \\
\ds \phantom{{\bf D}_{v_{\parallel}v_{\parallel}}} =
\frac{e^{2}}{m_{e}^{2}}\int_{0}^{t} dt''\int  d{\bf x}''\int d{\bf
v}''\; \langle{\bf\delta\bf E}_{\parallel}({\bf x},t){\cal G}({\bf x},{\bf
v},t;{\bf x}'',{\bf v}'',t''){\bf\delta\bf E}_{\parallel}({\bf
x},t'')\rangle.
\ea \label{eq:xr14}
\]
The other dif\/fusion coef\/f\/icients, ${\bf D}_{v_{\perp} v_{\perp}}$, 
${\bf D}_{v_{\perp} v_{\parallel}}$, ${\bf D}_{v_{\perp}\perp}$ can be
expressed in a similar way.

One can write a formal solution to the
Green's function equation as follows
\[
{\cal G}({\bf x},{\bf
v},t;{\bf x}',{\bf v}',t')=\Theta(t-t') {\bf\delta}({\bf x}-{\bf
x}'){\bf\delta}({\bf v}-{\bf v}')U(t,t'),
\label{eq:xr15}
\]
where the time propagator $U(t,t')$ is given by
\begin{equation}
U(t,t')=\exp{\left(-i\int_{t'}^{t}{\cal L}({\bf x},{\bf v},t'')\; dt''.
\right)}  \label{eq:r6}
\end{equation}
This allows us to simplify the expressions for the dif\/fusion coef\/f\/icients
\[
\tilde{\bf D}_{\perp\perp}             =  \int_{0}^{t} dt''\; \langle {\bf
v}_{E}({\bf x},t)U(t,t''){\bf v}_{E}({\bf
x},t'')\rangle,
\]
\[
{\bf D}_{\perp v_{\parallel}}          =  -\frac{e}{m_{e}}\int_{0}^{t}
dt''\; \langle {\bf v}_{E}({\bf x},t)U(t,t''){\bf\delta\bf E}_{\parallel}({\bf
x},t'')\rangle,
\]
\[
{\bf D}_{v_{\parallel} v_{\parallel}}  =
\frac{e^{2}}{m_{e}^{2}}\int_{0}^{t} dt''\; \langle{\bf\delta\bf
E}_{\parallel}({\bf x},t)U(t,t''){\bf\delta\bf E}_{\parallel}({\bf
x},t'')\rangle.  \label{eq:xr41}
\]

The solution to the equation (\ref{eq:r3}) can be formally written in terms of
the time propagator def\/ined through equation (\ref{eq:r6}), that is
\[
\delta f_{e}({\bf x},{\bf v},t)=U(t,0)\delta f_{e}({\bf
x},{\bf v},0)- i\int_{0}^{t}dt'\; U(t,t'){\cal L}_{1}\langle
f_{e}(\tilde{\bf x},v_{\perp},v_{\parallel},t')\rangle.   \label{eq:r7}
\]
This leads to the following Fourier component of the f\/luctuating part of
the distribution function
\begin{equation}
\delta f_{e\bf k\omega}({\bf v})=-i{\cal L}_{1{\bf k}\omega}\langle
f_{e}(\tilde{\bf x},v_{\perp},v_{\parallel},t)\rangle
\int_{t_{0}}^{t}dt'e^{i\omega(t-t')}\langle \; e^{-i{\bf k}\cdot{\bf x}}
U(t,t')e^{i{\bf k}\cdot{\bf x}}\rangle,   \label{eq:r8}
\end{equation}
where we have pulled out the velocity derivative of the averaged distribution
assuming that the time dependence of the average distribution function is much
slower than the time dependence of the orbits, and neglected the initial
condition.

Notice that the solution presented through equation (\ref{eq:r8}) has imbeded in it
two time scales. A fast time scale (associated with $\omega$, i.e., fast oscillation
time scale), and a slow time scale that should be related to the growth time scale.
The slow time scale is the classical quasi-linear time scale required for saturation.
The absence of a slow time scales leads to the absence of saturation.  The Fourier 
transform is clearly a transform over the fast time scale.

At this point we need to use a fundamental property of the time propagator,
which is
\[
\langle U(t,t')g({\bf x},{\bf v},t)\rangle=g\left({\bf x}'(t'),
{\bf v}'(t')\right)  \label{eq:r9}
\]
and
\[
\ba{l}
\ds U(t,t')  =  \exp{\left(-i\int_{t'}^{t}dt''\; {\cal L}(t'')\right)}
\vspace{3mm}\\
\ds \phantom{U(t,t')} =  \exp{\left(i\int_{t_{0}}^{t'}dt''\; {\cal
L}(t'')\right)} \exp{\left(-i\int_{t_{0}}^{t}dt''\; {\cal
L}(t'')\right)} =  U^{-1}(t',t_{0})U(t,t_{0}).
\ea   \label{eq:r10}
\]
Substituting this result into equation (\ref{eq:r8}) leads to the following
result
\[
\langle e^{-i{\bf k}\cdot{\bf x}}
U(t,t')e^{i{\bf k}\cdot{\bf x}}\rangle  =  \langle e^{-i{\bf k}\cdot
{\bf x}}U^{-1}(t',t_{0})U(t,t_{0})e^{i{\bf k}\cdot{\bf x}}\rangle
 =  \langle e^{-i{\bf k}\cdot{\bf
x}(t')}e^{i{\bf k}\cdot{\bf x}(t)}\rangle,  \label{eq:r11}
\]
where ${\bf x}(t)$ and ${\bf x}(t')$ represent the exact orbits of the
electrons. These orbits can be decomposed into the unperturbed orbits ${\bf
x}_{0}(t)$ plus the perturbation due to the nonlinear ef\/fects of the random
``bath'' of waves ${\bf\delta\bf x}(t)$, i.e.,
\[
{\bf x}(t)={\bf x}_{0}(t)+{\bf\delta\bf x}(t),
\label{eq:r12}
\]
where the perturbation can be expressed as follows
\[
{\bf\delta\bf x}(t)={\bf\delta\bf \tilde{x}}(t)+{\bf\delta\bf x}_{\parallel}+
\frac{{\bf\delta\bf v}_{\perp}}{\Omega_{e}}.
\label{eq:r12a}
\]

This f\/inally leads to the following expression for the perturbed distribution
function
\be
\ba{l}
\ds \delta f_{e\bf k\omega}({\bf v})  =  -i\frac{e}{m_{e}}\Phi_{\bf
k\omega}\left({\bf k}_{\perp}\cdot{\bf
v}_{\perp}\frac{1}{v_{\perp}}{\partial\over\partial
v_{\perp}}+k_{\parallel}{\partial\over\partial v_{\parallel}}+\frac{{\bf
k}\times\hat{\bf z}}{\Omega_{e}}\cdot\tilde{\bf\nabla}\right)\langle
f_{e}\rangle   \vspace{3mm} \\
\ds \qquad  \times  \int_{t_{0}}^{t}dt^{'} \; e^{\left(i\omega
(t-t')-i{\bf k}\cdot({\bf x}_{0}(t)-{\bf x}_{0}(t'))\right)}\langle
\exp{\left(i{\bf k}\cdot{\bf\Delta}{\bf x}(t,t')\right)}\rangle,
\ea
\label{eq:r13} 
\ee
where
\[
{\bf\Delta}{\bf x}(t,t')={\bf\delta\bf x}(t)-{\bf\delta\bf x}(t').
\label{eq:r14}
\]
At this point one can use the cumulant expansion (see for example
Weinstock~[20]) to write
\[
\langle\exp{\left(i{\bf k}\cdot{\bf\Delta}{\bf x}(t,t')\right)}
\rangle= \exp{\left(\sum_{n=1}^{\infty}\frac{1}{n!}\langle\left[i{\bf
k}\cdot{\bf\Delta\bf x}(t,t')\right]^{n}\rangle_{c}\right)},  \label{eq:r15}
\]
where $\langle\cdots\rangle_{c}$ is the cummulant, and note that this expansion
can be reduced to a single term in the case of a random variable of a Gaussian
process, that is
\begin{equation}
\langle\exp{\left(i{\bf k}\cdot{\bf\Delta}{\bf x}(t,t')\right)}\rangle=
\exp{\left(-\frac{1}{2}\langle\left[{\bf
k}\cdot{\bf\Delta\bf x}(t,t')\right]^{2}\rangle\right)}.
\label{eq:r16}
\end{equation}
The exponent can now be expanded in the following form
\be
\ba{l}
\ds
\langle\left[{\bf k}\cdot{\bf\Delta\bf x}(t,t')\right]^{2}\rangle  =
k_{\parallel}^{2}\left(\langle\delta x_{\parallel}^{2}(t)\rangle+
\langle\delta x_{\parallel}^{2}(t')\rangle-2\langle\delta
x_{\parallel}(t)\delta x_{\parallel}(t')\rangle\right)
\vspace{3mm} \\
\ds \qquad  +  {\bf
k}_{\perp}\cdot\left(\langle{\bf\delta\bf \tilde
{x}}_{\perp}^{2}(t)\rangle+\langle{\bf\delta\bf
\tilde{x}}_{\perp}^{2}(t')\rangle-2\langle{\bf\delta\bf
\tilde{x}}_{\perp}(t){\bf\delta\bf
\tilde{x}}_{\perp}(t')\rangle\right)\cdot{\bf k}_{\perp}
\vspace{3mm} \\
\ds \qquad
+  2 k_{\parallel}\left(\langle\delta x_{\parallel}(t){\bf\delta\bf
\tilde{x}}_{\perp}(t)\rangle+\langle\delta x_{\parallel}(t'){\bf\delta\bf
\tilde{x}}_{\perp}(t')\rangle-\langle\delta x_{\parallel}(t){\bf\delta\bf
\tilde{x}}_{\perp}(t')\rangle\right.
\vspace{3mm} \\
\ds \qquad  \left.
- \langle\delta x_{\parallel}(t'){\bf\delta\bf
\tilde{x}}_{\perp}(t)\rangle\right)\cdot{\bf k}_{\perp}
\vspace{3mm} \\
\ds \qquad  +  {\bf
k}_{\perp}\cdot\left(\frac{\langle{\bf\delta\bf
v}_{\perp}^{2}(t)\rangle}{\Omega_{e}^{2}}+\frac{\langle{\bf\delta\bf
v}_{\perp}^{2}(t')\rangle}{\Omega_{e}^{2}}-2\frac{\langle{\bf\delta\bf
v}_{\perp}(t){\bf\delta\bf
v}_{\perp}(t^{'})\rangle}{\Omega_{e}^{2}}\right)\cdot{\bf k}_{\perp}
\vspace{3mm} \\
\ds \qquad  +  2{\bf k}_{\perp}\cdot\left(\frac{\langle{\bf\delta\bf
\tilde{x}}_{\perp}(t){\bf\delta\bf
v}_{\perp}(t)\rangle}{\Omega_{e}}+\frac{\langle{\bf\delta\bf
\tilde{x}}_{\perp}(t'){\bf\delta\bf
v}_{\perp}(t')\rangle}{\Omega_{e}}-\frac{\langle{\bf\delta\bf
\tilde{x}}_{\perp}(t){\bf\delta\bf v}_{\perp}(t')\rangle}{\Omega_{e}}\right.
\vspace{3mm} \\
\ds \qquad  - \left.\frac{\langle{\bf\delta\bf
v}_{\perp}(t){\bf\delta\bf
\tilde{x}}_{\perp}(t')\rangle}{\Omega_{e}}\right)\cdot{\bf k}_{\perp}
\vspace{3mm} \\
\ds \qquad  +  2{\bf k}_{\perp}\cdot\left(\frac{\langle{\bf\delta\bf
v}_{\perp}(t)\delta x_{\parallel}(t)\rangle}{\Omega_{e}}+\frac{\langle
{\bf\delta\bf v}_{\perp}(t')\delta
x_{\parallel}(t')\rangle}{\Omega_{e}}-\frac{\langle{\bf\delta\bf
v}_{\perp}(t)\delta
x_{\parallel}(t')\rangle}{\Omega_{e}}\right.
\vspace{3mm} \\
\ds \qquad  - \left.\frac{\langle{\bf\delta\bf
v}_{\perp}(t')\delta
x_{\parallel}(t)\rangle}{\Omega_{e}}\right)k_{\parallel},
\ea
\label{eq:r17}     
\ee
where the f\/irst two terms and the fourth term on the right hand
side represent the parallel and perpendicular correlation functions,
while the other terms represent the cross correlation terms.

The dif\/fusion coef\/f\/icients are expressed in terms of the correlation function
of the corresponding random forces due to the background oscillations. They can
also be def\/ined more accurately as the rate of time change of velocity variance
around the mean value. The components of the velocity dif\/fusion tensor can be
written in the form:
\[
D_{v_{\alpha} v_{\beta}}=\frac{1}{2}\frac{d}{dt}\langle\delta v_{\alpha}(t)\delta
v_{\beta}(t)\rangle,  \label{eq:r18}
\]
where the subscripts $\alpha$ and $\beta$ represent the parallel and
perpendicular components. The mean value of the parallel velocity is zero,
while the mean value of the perpendicular velocity is given by the ${\bf
E}_{0}\times{\bf B}$ drift due to the electric f\/ield of the electrojet.

The random velocity components on the other hand are given by
\[
\ba{l}
\ds \delta v_{\parallel}(t)  =  -\frac{e}{m_{e}}\int_{0}^{t}dt'\;
\delta E_{\parallel} ({\bf x}(t'),t'),
\vspace{3mm}\\
\ds {\bf\delta\bf v}_{\perp}(t)  =  -\frac{c}{B}
{\bf\delta\bf E}({\bf x}(t),t)\times\hat{\bf z},   \label{eq:r19}
\ea
\]
where we have set the initial time to be $t_{0}=0$. We then have an expression
for the guiding center spatial displacements
\[
\ba{l}
\ds \delta x_{\parallel}(t)      =  -\frac{e}{m_{e}}\int_{0}^{t}dt_{1}
\int_{0}^{t_{1}}dt_{2}\; \delta E_{\parallel} ({\bf x}(t_{2}),t_{2}),
 \vspace{3mm}\\
\ds {\bf\delta\bf\tilde{x}}_{\perp}(t)  =  -\frac{c}{B}\int_{0}^{t}dt_{1}\;
({\bf\delta\bf E}({\bf x}(t_{1}),t_{1})\times\hat{\bf z}).
\ea
\label{eq:r20}
\]

\noindent
{\bfseries \itshape Parallel Velocity Diffusion.}
Following the work of Ishihara {\it et al.}~[12], the parallel dif\/fusion
coef\/f\/icient $D_{v_{\parallel}v_{\parallel}}$ can be obtained by evaluating the
parallel velocity correlation function
\[
\ba{l}
\ds \langle\delta v_{\parallel}^{2}(t)\rangle  =  \frac{e^{2}}{m_{e}^{2}}
\int_{0}^{t}dt'\int_{0}^{t}dt''\; \langle\delta E_{\parallel}({\bf
x}(t'),t')\delta E_{\parallel}({\bf x}(t''),t'')\rangle
 \vspace{3mm}\\
\ds \phantom{\langle\delta v_{\parallel}^{2}(t)\rangle}
  =  \frac{e^{2}}{m_{e}^{2}}
\int_{0}^{t}ds\int_{s-t}^{s}ds'\; \langle\delta E_{\parallel}({\bf
x}(s),s)\delta E_{\parallel}({\bf x}(s-s'),s-s')\rangle.
\ea
 \label{eq:r21}
\]
Using the def\/inition of the exact orbits leads to
\[
\langle\delta v_{\parallel}^{2}(t)\rangle=\frac{e^{2}}{m_{e}^{2}}
\int_{0}^{t}ds\int_{s-t}^{s}ds'\sum_{\bf k}|\delta E_{\parallel\bf k}|^{2}
e^{i\left({\bf k}\cdot({\bf x}_{0}(s)-{\bf x}_{0}(s-s'))-\omega_{\bf k}s'
\right)}\langle e^{i{\bf k}\cdot{\bf\Delta\bf x}(s,s-s')}\rangle.
\label{eq:r22}
\]
It is clear from this equation that one needs the correlation of the spatial
displacements to evaluate the parallel as well as the perpendicular and cross
dif\/fusion coef\/f\/icients. In general, we need three correlations functions:
$\langle\delta x_{\parallel}(\tau)\delta x_{\parallel}(\tau')\rangle$,
$\langle\delta x_{\parallel}(\tau){\bf\delta\bf x}_{\perp}(\tau')\rangle$,
and $\langle{\bf\delta\bf x}_{\perp}(\tau){\bf\delta\bf x}_{\perp}(\tau')
\rangle$. The f\/irst correlation function is given by
\[
\langle\delta x_{\parallel}(\tau)\delta x_{\parallel}(\tau')\rangle=
\frac{e^{2}}{m_{e}^{2}}\int_{0}^{\tau}dt_{1}\int_{0}^{t_{1}}dt_{2}\int_{0}^{
\tau'}dt_{1}'\int_{0}^{t_{1}'}dt_{2}'\; \langle\delta
E_{\parallel}({\bf x}( t_{2}),t_{2})\delta E_{\parallel}({\bf
x}(t_{2}'),t_{2}')\rangle  \label{eq:r23}
\]
following the arguments of Ishihara {\it et al.}~[12],
one obtains the following
result for the parallel correlation function
\[
\langle\delta x_{\parallel}(\tau)\delta x_{\parallel}(\tau')\rangle  =
\left\{
\ba{ll}
\ds \frac{1}{3}D_{\parallel\parallel}{\tau'}^2(3\tau-\tau') & \mbox{for}
\quad \tau\geq\tau', \vspace{3mm} \\
\ds \frac{1}{3}D_{\parallel\parallel}\tau^{2}(3\tau'-\tau) &
\mbox{for} \quad \tau'\geq\tau.
\ea \right.
\label{eq:r24}
\]

\noindent
{\bfseries \itshape Perpendicular Guiding Center Spatial Diffusion.}
We now proceed to evaluate the perpendicular correlation function
\[
\langle{\bf\delta\bf \tilde{x}}_{\perp}(\tau){\bf\delta\bf
\tilde{x}}_{\perp}(\tau')\rangle=\frac{c^{2}}{B^{2}}
\int_{0}^{\tau}dt_{1}\int_{0}^{ \tau'}dt_{2}\; \langle({\bf\delta\bf
E}({\bf x}( t_{1}),t_{1})\times\hat{\bf z}){\bf\delta\bf E}({\bf
x}(t_{2},t_{2})\times \hat{\bf z})\rangle.
\label{eq:r25}
\]
It is important to note that this correlation function was approximated by Dum
and Dupree~[7] by
\[
\langle{\bf\delta\bf \tilde{x}}_{\perp}(\tau){\bf\delta\bf
\tilde{x}}_{\perp}(\tau')\rangle_{\mbox{\scriptsize Dum-Dupree}}\propto
\tilde{\bf D}_{\bf\perp\bf\perp}(\tau-\tau')
\label{eq:r26}
\]
using a Markovian approximation, and assuming the dif\/fusion coef\/f\/icient to be
independent of time. However, such an approximation has its physical limitations
as shown by Salat~[18] and Ishihara {\it et al.}~[12].

We now proceed to evaluate the perpendicular correlation function by noting
that by def\/inition the dif\/fusion coef\/f\/icient describing the dif\/fusion of the
electron guiding centers is given 
\be
\ba{l}
\ds \tilde{\bf D}_{\bf\perp\bf\perp}(t)  =
\frac{1}{2}\frac{d}{dt}\langle{\bf\delta\bf \tilde{x}}_{\perp}
(t){\bf\delta\bf 
\tilde{x}}_{\perp}(t)\rangle
\vspace{3mm} \\
\ds \phantom{\tilde{\bf D}_{\bf\perp\bf\perp}(t) }  =
\frac{1}{2}\frac{c^{2}}{B^{2}}\frac{d}{dt}\int_{0}^{t}dt_{1}\int_{0}^{
t}dt_{2}\; \langle({\bf\delta\bf E}({\bf x}(t_{1}),t_{1})\times\hat{\bf
z})({\bf\delta\bf E}({\bf x}(t_{2}),t_{2})\times \hat{\bf z})\rangle
\vspace{3mm} \\
\ds \phantom{\tilde{\bf D}_{\bf\perp\bf\perp}(t) }
  =  \frac{c^{2}}{B^{2}}\int_{0}^{t}dt_{1}\;
\langle({\bf\delta\bf E}({\bf x}(t_{1}),t_{1})\times\hat{\bf
z}){\bf\delta\bf E}({\bf x}(t),t)\times \hat{\bf z})\rangle.
\ea
\label{eq:r27} 
\ee
This in turn leads, in the case $\tau\geq\tau'$, to
\[
\ba{l}
\ds \langle{\bf\delta\bf \tilde{x}}_{\perp}(\tau){\bf\delta\bf
\tilde{x}}_{\perp}(\tau')\rangle  =
\frac{c^{2}}{B^{2}}\int_{0}^{\tau'}dt_{1}
\int_{0}^{\tau'}dt_{2}\; \langle({\bf\delta\bf E}({\bf
x}(t_{1}),t_{1})\times\hat{\bf z})({\bf\delta\bf E}({\bf x}(t_{2}),t_{2})\times
\hat{\bf z})\rangle
\vspace{3mm}\\
\ds \qquad  + \frac{c^{2}}{B^{2}}\int_{\tau'}^{\tau}
dt_{1}\int_{0}^{\tau'}dt_{2}\; \langle({\bf\delta\bf E}({\bf
x}(t_{1}),t_{1})\times\hat{\bf z})({\bf\delta\bf E}({\bf x}(t_{2}),t_{2})\times
\hat{\bf z})\rangle.
\ea
\label{eq:r28}
\]
The f\/irst term can be
approximated using the def\/inition of the dif\/fusion coef\/f\/icient 
\[
\ba{l}
\ds \langle{\bf\delta\bf \tilde{x}}_{\perp}(\tau)
{\bf\delta\bf \tilde{x}}_{\perp}(\tau')\rangle  =  2\tilde{\bf
D}_{\bf\perp\bf\perp}(\tau') \tau'
\vspace{3mm} \\
\ds \qquad  +  \frac{c^{2}}{B^{2}}
\int_{\tau'}^{\tau}dt_{1}\int_{0}^{\tau'}dt_{2}\; \langle({\bf\delta\bf
E}({\bf x}(t_{1}),t_{1})\times\hat{\bf z})({\bf\delta\bf E}({\bf
x}(t_{2}),t_{2})\times \hat{\bf z})\rangle.
\ea
\label{eq:r29}
\]
Using the def\/inition of the dif\/fusion coef\/f\/icient given by equation
(\ref{eq:r27}) and changing variables allows us to rewrite the integral in the
following form
\be
\ba{l}
\ds \langle{\bf\delta\bf \tilde{x}}_{\perp}(\tau)
{\bf\delta\bf \tilde{x}}_{\perp}(\tau')\rangle  =  2\tilde{\bf
D}_{\bf\perp\bf\perp}(\tau')
\tau'+\frac{c^{2}}{B^{2}}\int_{0}^{\tau-\tau'}d\xi\int_{0}^{\tau'}
d\eta \vspace{3mm} \\
\ds \qquad \times  \langle({\bf\delta\bf E}({\bf
x}(\xi+\tau'),\xi+\tau')\times\hat{\bf z})({\bf\delta\bf E}({\bf
x}(\eta),\eta)\times \hat{\bf z})\rangle.
\ea
\label{eq:r30}
\ee

At this stage we would like to consider the case where $\tau\approx\tau'$,
i.e., we shall consider the case where $|\tau-\tau'|/\tau'$ is a small
parameter. This assumption is critical to the rest of the calculations in this
paper. It is also important to note that in general $\tau$ and $\tau'$ are
independent variables, which makes our assumption a critical one indeed.
We then can write
\[
{\bf\delta\bf E}\left({\bf x}(\xi+\tau'),\xi+\tau'\right)={\bf\delta\bf
E}({\bf x} (\tau'),\tau')+\xi\frac{d}{d\tau'}{\bf\delta\bf
E}({\bf x}(\tau'),\tau'), \label{eq:r31}
\]
where the convective derivative is def\/ined by
\[
\frac{d}{d\tau'}=\left({\partial\over\partial\tau'}+{\bf
v}(\tau')\cdot{\bf\nabla}\right).
\label{eq:r32}
\]
The f\/irst term when substituted into equation (\ref{eq:r30}) leads to
$\tilde{\bf D}_{\bf\perp\bf\perp}(\tau')(\tau-\tau')$ and therefore the
expression becomes 
\be
\ba{l}
\ds \langle{\bf\delta\bf \tilde{x}}_{\perp}(\tau)
{\bf\delta\bf \tilde{x}}_{\perp}(\tau')\rangle  =
\tilde{\bf D}_{\bf\perp\bf\perp}(\tau') (\tau+\tau')
\vspace{3mm}\\
\ds \qquad +\frac{c^{2}}{B^{2}}
\int_{0}^{\tau-\tau'}\xi \; d\xi\int_{0}^{\tau'}d\eta\; 
\langle\left(
\frac{d}{d\tau'}{\bf\delta\bf E}({\bf x}(\tau'),\tau')
\times\hat{\bf z}\right)({\bf\delta\bf E}({\bf x}(\eta),\eta)
\times \hat{\bf z})\rangle.
\ea
\label{eq:r33}
\ee
We now use the following property of the dif\/fusion coef\/f\/icient
\be
\ba{l}
\ds \frac{d}{dt}\tilde{\bf D}_{\bf\perp\bf\perp}(t)  =
\frac{c^{2}}{B^{2}}\langle\left({\bf\delta\bf E}({\bf x}(t),t)
\times\hat{\bf z}\right)\left({\bf\delta\bf E}({\bf x}(t),t)\times
\hat{\bf z}\right)\rangle
\vspace{3mm} \\
\ds \qquad  +  \frac{c^{2}}{B^{2}}
\int_{0}^{t}dt_{1}\; \langle\left(\frac{d}{dt}{\bf\delta\bf
E}({\bf x}(t),t)\times\hat{\bf z}\right)\left({\bf\delta\bf E}({\bf
x}(t_{1}),t_{1})\times \hat{\bf z}\right)\rangle.
\ea
\label{eq:r34}
\ee
Consequently we can express the two-time correlation function by substituting
expression~(\ref{eq:r34}) into expression (\ref{eq:r33})
\[
\ba{l}
\ds \langle{\bf\delta\bf \tilde{x}}_{\perp}(\tau)
{\bf\delta\bf \tilde{x}}_{\perp}(\tau')\rangle  =  \tilde{\bf
D}_{\bf\perp\bf\perp}(\tau') (\tau+\tau')+\frac{(\tau-\tau')^{2}}{2}
\left(\frac{d}{d\tau'}\tilde{\bf D}_{\bf\perp\bf\perp}(\tau')\right.
\vspace{3mm} \\
\ds \qquad -  \left.
\frac{c^{2}}{B^{2}}\langle({\bf\delta\bf E}({\bf x}(\tau'),\tau')
\times\hat{\bf z})({\bf\delta\bf E}({\bf x}(\tau'),\tau')\times\hat{\bf
z})\rangle\right).
\ea
\label{eq:r35}
\]
Therefore
\be
\ba{l}
\langle{\bf\delta\bf \tilde{x}}_{\perp}(\tau)
{\bf\delta\bf \tilde{x}}_{\perp}(\tau)\rangle+\langle{\bf\delta\bf
\tilde{x}}_{\perp}(\tau'){\bf\delta\bf \tilde{x}}_{\perp}(\tau')\rangle-
2\langle{\bf\delta\bf \tilde{x}}_{\perp}(\tau)
{\bf\delta\bf \tilde{x}}_{\perp}(\tau')\rangle
\vspace{3mm}\\
\ds \qquad = 2\tilde{\bf
D}_{\bf\perp\bf\perp}(\tau)\tau+2\tilde{\bf
D}_{\bf\perp\bf\perp}(\tau')\tau'-2\tilde{\bf
D}_{\bf\perp\bf\perp}(\tau')(\tau+\tau')
\vspace{3mm} \\
\ds \qquad  -(\tau-\tau')^{2}
\left(\frac{d}{d\tau'}\tilde{\bf D}_{\bf\perp\bf\perp}(\tau')
-\frac{c^{2}}{B^{2}}\langle({\bf\delta\bf E}({\bf x}(\tau'),\tau')
\times\hat{\bf z})({\bf\delta\bf E}({\bf x}(\tau'),
\tau')\times\hat{\bf z}) \rangle\right).
\ea
\label{eq:r36}
\ee
The f\/irst four terms on the right hand side of expression (\ref{eq:r36}) can
be rewritten, after expanding $\tilde{\bf D}_{\bf\perp\bf\perp}(\tau)$ around
$\tau'$, in the following form,
\be
2\left(\tilde{\bf D}_{\bf\perp\bf\perp}(\tau)-\tilde{\bf
D}_{\bf\perp\bf\perp}(\tau')\right)\tau
-(\tau-\tau')^{2}\frac{d}{d\tau'}\tilde{\bf
D}_{\bf\perp\bf\perp}(\tau')=
(\tau^{2}-{\tau'}2)\frac{d}{d\tau'}\tilde{\bf D}_{\bf\perp\bf\perp}
(\tau'),
\label{eq:r37}
\end{equation}
where we have used
\[
\tilde{\bf D}_{\bf\perp\bf\perp}(\tau)=\tilde{\bf
D}_{\bf\perp\bf\perp}(\tau')+(\tau-\tau')
\frac{d}{d\tau'}\tilde{\bf D}_{\bf\perp\bf\perp}(\tau').
\label{eq:r38}
\]
Finally equation (\ref{eq:r36}) can be written as
\be
\ba{l}
\ds \langle{\bf\delta\bf \tilde{x}}_{\perp}(\tau)
{\bf\delta\bf \tilde{x}}_{\perp}(\tau)\rangle+\langle{\bf\delta\bf
\tilde{x}}_{\perp}(\tau'){\bf\delta\bf \tilde{x}}_{\perp}(\tau')\rangle-
2\langle{\bf\delta\bf \tilde{x}}_{\perp}(\tau)
{\bf\delta\bf \tilde{x}}_{\perp}(\tau')\rangle
 =  (\tau^{2}-{\tau'}^2)\frac{d}{d\tau'}\tilde{\bf
D}_{\bf\perp\bf\perp}(\tau')
\vspace{3mm}\\
\ds \qquad +(\tau-\tau')^{2}\frac{c^{2}}{B^{2}}
\langle({\bf\delta \bf E}({\bf
x}(\tau'), \tau')\times\hat{\bf z})({\bf\delta\bf E}({\bf
x}(\tau'),\tau')\times\hat{\bf z}) \rangle.
\ea\hspace{-19.8pt}
\label{eq:r39}
\ee
The second term can easily be identif\/ied with $\langle{\bf\delta\bf
v}_{\perp}(\tau'){\bf\delta\bf v}_{\perp}(\tau')\rangle$, and can
therefore be identif\/ied with a velocity space dif\/fusion coef\/f\/icient, which
decribes random changes of the gyroradius and phase angle, and can therefore
be written as 
\begin{equation} 
\langle{\bf\delta\bf
v}_{\perp}(\tau'){\bf\delta\bf
v}_{\perp}(\tau')\rangle=2D_{v_{\perp}v_{\perp}}(\tau')\tau'.
\label{eq:r40} 
\end{equation}
Substituting expression (\ref{eq:r40}) into equation (\ref{eq:r39}) leads
to
\be
\ba{l}
\ds \langle{\bf\delta\bf \tilde{x}}_{\perp}(\tau)
{\bf\delta\bf \tilde{x}}_{\perp}(\tau)\rangle+\langle{\bf\delta\bf
\tilde{x}}_{\perp}(\tau'){\bf\delta\bf \tilde{x}}_{\perp}(\tau')\rangle-
2\langle{\bf\delta\bf \tilde{x}}_{\perp}(\tau)
{\bf\delta\bf \tilde{x}}_{\perp}(\tau')\rangle
 \vspace{3mm} \\
\ds \qquad = (\tau^{2}-{\tau'}^2)
\frac{d}{d\tau'}\tilde{\bf D}_{\bf\perp\bf\perp}(\tau')+2\Omega_{e}^{2}
(\tau-\tau')^{2}\tau'\left(\frac{D_{v_{\perp}v_{\perp}}(\tau')}{
\Omega_{e}^{2}}\right).
\ea
\label{eq:r41}
\ee
One can f\/inally use the approximation of Dum and Dupree~[7], that is the
velocity dif\/fusion makes approximately an equal contribution to the dif\/fusion
of the guiding centers, i.e.,
\begin{equation}
\left(\frac{D_{v_{\perp}v_{\perp}}(\tau')}{\Omega_{e}^{2}}\right)\approx
\tilde{\bf D}_{\bf\perp\bf\perp}(\tau')  \label{eq:r42}
\end{equation}
which f\/inally leads to the expression for the perpendicular correlation
functions (\ref{eq:r41})
\be
\ba{l}
\ds \langle{\bf\delta\bf \tilde{x}}_{\perp}(\tau)
{\bf\delta\bf \tilde{x}}_{\perp}(\tau)\rangle+\langle{\bf\delta\bf
\tilde{x}}_{\perp}(\tau'){\bf\delta\bf \tilde{x}}_{\perp}(\tau')\rangle-
2\langle{\bf\delta\bf \tilde{x}}_{\perp}(\tau)
{\bf\delta\bf \tilde{x}}_{\perp}(\tau')\rangle
 \vspace{3mm} \\
\ds \qquad = (\tau^{2}-{\tau'}^2)
\frac{d}{d\tau'}\tilde{\bf D}_{\bf\perp\bf\perp}(\tau')+2\Omega_{e}^{2}
(\tau-\tau')^{2}\tau'\tilde{\bf D}_{\bf\perp\bf\perp}(\tau').
\ea
\label{eq:r43}
\ee
In the case $\tau'\geq\tau$ we just exchange the role of $\tau$ and
$\tau'$ in the expression (\ref{eq:r43}).

\noindent
{\bfseries \itshape Perpendicular Velocity Diffusion.}
For the velocity dif\/fusion, with $\tau\geq\tau'$ we need to evaluate the
following expression
\begin{equation}
\langle{\bf\delta\bf v}_{\perp}(\tau){\bf\delta\bf v}_{\perp}(\tau')
\rangle \simeq\langle{\bf\delta\bf v}_{\perp}(\tau'){\bf\delta\bf
v}_{\perp}(\tau')\rangle +\frac{(\tau-\tau')}{2}{\partial\over\partial
\tau'}\langle{\bf\delta \bf v}_{\perp}(\tau'){\bf\delta\bf v}_{\perp}
(\tau')\rangle.   \label{eq:r43a}
\end{equation}
Using the def\/inition of the velocity dif\/fusion we can express the right hand
side of equation~(\ref{eq:r43a}) as follows
\[
\langle{\bf\delta\bf v}_{\perp}(\tau){\bf\delta\bf v}_{\perp}(\tau')\rangle
\simeq 2{\bf D}_{v_{\perp}v_{\perp}}(\tau')\tau'+(\tau'-\tau){\partial
({\bf D}_{v_{\perp}v_{\perp}}(\tau')\tau')\over\partial\tau'}
\label{eq:r43b} 
\]
which f\/inally leads to the contribution of perpendicular velocity dif\/fusion
\[
\ba{l}
\ds
\langle{\bf\delta\bf v}_{\perp}(\tau){\bf\delta\bf
v}_{\perp}(\tau)\rangle+\langle{\bf\delta\bf v}_{\perp}(\tau'){\bf\delta\bf
v}_{\perp}(\tau')\rangle-2\langle{\bf\delta\bf v}_{\perp}(\tau){\bf
\delta\bf v}_{\perp}(\tau')\rangle
\vspace{3mm}\\
\ds \qquad\!\simeq
 2{\bf D}_{v_{\perp}v_{\perp}}(\tau)\tau-2{\bf
D}_{v_{\perp}v_{\perp}}(\tau') \tau'-2(\tau-\tau'){\bf
D}_{v_{\perp}v_{\perp}}(\tau')-2\tau'(\tau-\tau'){\partial {\bf
D}_{v_{\perp}v_{\perp}}(\tau')\over\partial\tau'}.
\ea
\label{eq:r43c}
\]
Expanding the f\/irst term on the right hand side leads to
\[
{\bf D}_{v_{\perp}v_{\perp}}(\tau)\simeq{\bf D}_{v_{\perp}v_{\perp}}
(\tau')+ (\tau-\tau'){\partial {\bf D}_{v_{\perp}v_{\perp}}
(\tau')\over\partial \tau'}  \label{eq:r43d}
\]
which leads to
\[
\frac{\langle{\bf\delta\bf v}_{\perp}(\tau){\bf\delta\bf
v}_{\perp}(\tau)\rangle}{\Omega_{e}^{2}}+\frac{\langle{\bf\delta\bf
v}_{\perp}(\tau'){\bf\delta\bf
v}_{\perp}(\tau')\rangle}{\Omega_{e}^{2}}-2\frac{\langle{\bf\delta\bf
v}_{\perp}(\tau){\bf \delta\bf
v}_{\perp}(\tau')\rangle}{\Omega_{e}^{2}}\simeq 2(\tau-\tau')^{2}
{\partial \tilde{\bf D}_{\perp\perp}(\tau')\over\partial \tau'},
\label{eq:r43e} 
\]
where we have used equation (\ref{eq:r42}).

\noindent
{\bfseries \itshape Parallel Velocity-Perpendicular Spatial Cross Diffusion.}
We have now evaluated the parallel and perpendicular correlation functions,
we therefore only need to calculate the cross correlation, which is given by
\begin{equation} 
\!\!\langle\delta x_{\parallel}(\tau)
{\bf\delta\bf \tilde{x}}_{\perp}(\tau')\rangle=\frac{e}{m_{e}}\frac{c}{B}
\int_{0}^{\tau}\!dt_{1}\!\int_{0}^{t_{1}}\!dt_{2}\!\int_{0}^{\tau'}\!\!dt_{1}'
\;\langle\delta E_{\parallel}({\bf x}(t_{2}),t_{2}){\bf\delta\bf E}
({\bf x}( t_{1}'),t_{1}')\times\hat{\bf
z})\rangle.   \label{eq:r44}
\end{equation}

We proceed in a similar fashion to the previous cases, and start by def\/ining
the cross dif\/fusion coef\/f\/icient
\[
\ba{l}
\ds \underline{D}_{v_{\parallel}\perp}(t)  =
 \frac{1}{2}\frac{d}{dt}\langle\delta
v_{\parallel}(t){\bf\delta\bf \tilde{x}}_{\perp}(t)\rangle
\vspace{3mm}\\
\ds \phantom{\underline{D}_{v_{\parallel}\perp}(t)} =
\frac{1}{2}\frac{e}{m_{e}}\frac{c}{B} \int_{0}^{t}dt_{1}\; 
\langle\delta E_{\parallel}({\bf x}(
t_{1}),t_{1}){\bf\delta\bf E}({\bf x}(t),t)\times\hat{\bf
z})\rangle
\vspace{3mm} \\
\ds \phantom{\underline{D}_{v_{\parallel}\perp}(t)} +
\frac{1}{2}\frac{e}{m_{e}}\frac{c}{B} 
\int_{0}^{t}dt_{1}\; \langle\delta E_{\parallel}({\bf x}(
t),t){\bf\delta\bf E}({\bf x}(t),t)\times\hat{\bf z})\rangle.
\ea
\label{eq:r45}
\]
In the case $\tau\geq\tau'$ we can decompose expression (\ref{eq:r44})
and write it in the following form
\[
\langle\delta v_{\parallel}(t){\bf\delta\bf x}_{\perp}(t)\rangle=
\frac{1}{2}\frac{e}{m_{e}}\frac{c}{B}\int_{0}^{t}dt_{1}\int_{0}^{t}dt_{2}
\; \langle\delta E_{\parallel}({\bf x}(t_{1}),t_{1}){\bf\delta\bf E}({\bf
x}(t_{2}),t_{2})\times\hat{\bf z})\rangle  \label{eq:r46}
\]
which allows us to approximate the integrant of equation (\ref{eq:r44}) as
follows
\[
\ba{l}
\ds \frac{e}{m_{e}}\frac{c}{B}\int_{0}^{t_{1}}dt_{2}\int_{0}^{
\tau'}dt_{1}'\; \langle\delta E_{\parallel}({\bf x}(
t_{2}),t_{2}){\bf\delta\bf E}({\bf x}( t_{1}'),t_{1}')\times\hat{\bf
z})\rangle
\vspace{3mm}\\
\ds \qquad \approx
 \frac{e}{m_{e}}\frac{c}{B}\int_{0}^{\min(t_{1},\tau')}dt_{1}\int_{0}^{
\min(t_{1},\tau')}dt_{1}'\; \langle\delta E_{\parallel}({\bf x}(
t_{2}),t_{2}){\bf\delta\bf E}({\bf x}( t_{1}'),t_{1}')\times\hat{\bf
z})\rangle.
\ea \label{eq:r47}
\]
The right hand side of this equation can easily be identif\/ied as
\[
\langle\delta v_{\parallel}(t){\bf\delta\bf x}_{\perp}(t)\rangle_{t=
\min(t_{1},\tau')}\approx
2\underline{D}_{v_{\parallel}\perp}\min(t_{1},\tau').
\label{eq:r48}
\]
Expression (\ref{eq:r44}) can now be approximated as
\[
\langle\delta x_{\parallel}(\tau)
{\bf\delta\bf \tilde{x}}_{\perp}(\tau')\rangle=\int_{0}^{\tau'}dt_{1}\;
2\underline{D}_{v_{\parallel}\perp}t_{1}+\int_{\tau'}^{\tau}dt_{1}\; 2
\underline{D}_{v_{\parallel}\perp}
\tau'=\underline{D}_{v_{\parallel}\perp}(2\tau-\tau')\tau'.
\label{eq:r49} 
\]
On the other hand
\[
\langle\delta x_{\parallel}(\tau')
{\bf\delta\bf \tilde{x}}_{\perp}(\tau)\rangle=\frac{e}{m_{e}}\frac{c}{B}
\int_{0}^{\tau'}dt_{1}\int_{0}^{t_{1}}dt_{2}\int_{0}^{
\tau}dt_{1}'\; \langle\delta E_{\parallel}({\bf x}(
t_{2}),t_{2})\left({\bf\delta\bf E}({\bf x}( t_{1}'),
t_{1}')\times\hat{\bf z}\right)\rangle.   \label{eq:r50}
\]
This expression can be evaluated for $\tau\geq\tau'$ following the same
procedure as before to obtain
\[
\langle\delta x_{\parallel}(\tau')
{\bf\delta\bf \tilde{x}}_{\perp}(\tau)\rangle=\int_{0}^{\tau'}dt_{1}\;
2\underline{D}_{v_{\parallel}\perp}\min(t_{1},\tau)
=\underline{D}_{v_{\parallel} \perp}{\tau'}^2.   \label{eq:r51}
\]
This f\/inally allows to sum all the cross correlation functions that appear in
the expression~(\ref{eq:r17})
\[
\ba{l}
\ds \langle\delta x_{\parallel}(t){\bf\delta\bf \tilde{x}}_{\perp}(t)
\rangle+\langle\delta x_{\parallel}(t'){\bf\delta\bf
\tilde{x}}_{\perp}(t')\rangle
\vspace{3mm}\\
\ds \qquad  =  \langle\delta
x_{\parallel}(t){\bf\delta\bf \tilde{x}}_{\perp}(t')\rangle+\langle\delta
x_{\parallel}(t'){\bf\delta\bf \tilde{x}}_{\perp}(t)\rangle
 +\underline{D}_{v_{\parallel}\perp}(t')(t-t')^{2}+ (t-t')t^{2}
 {\partial \underline{D}_{v_{\parallel}\perp}\over\partial t'}.
\ea
\label{eq:r52}
\]

\noindent{\bfseries \itshape Cross Perpendicular Velocity-Space Diffusion.}
One can easily show that the cross perpendicular dif\/fusion (velocity-space)
can be written as follows
\[
\ba{l}
\ds \langle\delta v_{\perp}(\tau){\bf\delta\bf\tilde{x}
}_{\perp}(\tau)\rangle+\langle\delta v_{\perp}(\tau'){\bf\delta\bf
\tilde{x}}_{\perp}(\tau')\rangle
\vspace{3mm}\\
\ds \qquad =  \langle\delta
v_{\perp}(\tau){\bf\delta\bf \tilde{x}}_{\perp}(\tau')\rangle+
\langle\delta v_{\perp}(\tau'){\bf\delta\bf\tilde{x}}_{\perp}
(\tau)\rangle  +  2(\tau-\tau^{'})^{2}{\partial
{\bf D}_{v_{\perp}\perp}(\tau')\over \partial\tau'}.
\ea  \label{eq:r52b}
\]

\noindent
{\bfseries \itshape Cross Velocity diffusion}
\[
\ba{l}
\ds \langle\delta x_{\parallel}(\tau){\bf\delta\bf v}_{\perp}(\tau')
\rangle  =
\underline{D}_{v_{\parallel}v_{\perp}}(\tau')\tau'(2\tau-\tau'),
\vspace{3mm} \\
\langle\delta x_{\parallel}(\tau'){\bf\delta\bf v}_{\perp}(\tau)
\rangle  =  \underline{D}_{v_{\parallel}v_{\perp}}(\tau')\tau^{'2}.
\ea
\label{eq:r52x} 
\]
which f\/inally leads to the following expression
\[
\ba{l}
\ds \langle\delta x_{\parallel}(t){\bf\delta\bf v}_{\perp}(t)
\rangle+\langle\delta x_{\parallel}(t'){\bf\delta\bf
v}_{\perp}(t')\rangle
\vspace{3mm}\\
\ds \qquad  =  \langle\delta
x_{\parallel}(t){\bf\delta\bf v}_{\perp}(t')\rangle+\langle\delta
x_{\parallel}(t'){\bf\delta\bf v}_{\perp}(t)\rangle +
\underline{D}_{v_{\parallel}v_{\perp}}(t')(t-t')^{2}+
(t-t')t^{2}{\partial\underline{D}_{v_{\parallel}v_{\perp}}
\over\partial t'}.
\ea
\label{eq:r52y}  
\]

We now are able to express equation (\ref{eq:r37}) in terms of components
of the dif\/fusion tensor, that is for $t\geq t'$
\be
\ba{l}
\ds \langle\left[{\bf k}\cdot{\bf\Delta\bf x}(t,t')\right]^{2}\rangle  =
k_{\parallel}^{2}\left(\frac{2}{3}D_{v_{\parallel}v_{\parallel}}(t^{3}
+{t'}^2(2t'-3t))\right) +  {\bf k}_{\perp}\cdot\left(\left[(t^{2}-{t'}^2)
\frac{d}{dt'}\tilde{\bf
D}_{\bf\perp\bf\perp}(t')\right.\right.  \vspace{3mm}\\
\ds \qquad  + \left.\left. 2(t-t')^{2}{\partial\over \partial t'}
\left\{ \tilde{\bf D}_{\bf\perp\bf\perp}(t')+
\frac{{\bf D}_{v_{\perp}\perp}(t')}
{\Omega_{e}}\right\}\right]
 +  2\Omega_{e}^{2} (t-t')^{2}t'\tilde{\bf D}_{\bf\perp\bf\perp}(t')
\right)\cdot{\bf k}_{\perp}
\vspace{3mm}\\
\ds \qquad  +  2 k_{\parallel}\left(\left[\underline{D}_{v_{\parallel}
\bf\perp}(t')+\frac{\underline{D}_{v_{\parallel}v_{\perp}}(t')}{\Omega_{e}}
\right](t-t')^{2} \right.
\vspace{3mm}\\
\ds \qquad \left. + t^{2}(t-t'){\partial\over\partial t'}\left[
\underline{D}_{v_{\parallel}\perp}(t')
+\frac{\underline{D}_{v_{\parallel}
v_{\perp}}(t')}{\Omega_{e}}\right]\right)\cdot
{\bf k}_{\perp}.
\ea \hspace{-22.1pt}
  \label{eq:r53}
\ee

At this stage we need to solve self-consistently for the dif\/fusion
coef\/f\/icients introduced above. First, we start with the parallel
dif\/fusion 
\be
\ba{l}
\ds D_{v_{\parallel}v_{\parallel}}  =  \frac{1}{2}\frac{d}{dt}\langle\delta
v_{\parallel}(t)\delta v_{\parallel}(t)\rangle  
\vspace{3mm}\\
\ds  
\qquad =\frac{1}{2}\frac{d}{dt}\frac{e^{2}}{m_{e}^{2}}
\int_{0}^{t}ds\int_{s-t}^{s}ds'\sum_{\bf k}|\delta E_{\parallel\bf k}|^{2}
e^{i\left({\bf k}\cdot({\bf x}_{0}(s)-{\bf x}_{0}(s-s'))-
\omega_{\bf k}s'\right)}\langle e^{i{\bf k}\cdot{\bf\Delta\bf x}
(s,s-s')}\rangle  
\vspace{3mm}\\
\ds \qquad = 
\frac{1}{2}\frac{d}{dt}\frac{e^{2}}{m_{e}^{2}}
\int_{0}^{t}ds\int_{s-t}^{0}ds'\sum_{\bf k}|\delta E_{\parallel\bf k}|^{2}
e^{i\left({\bf k}\cdot({\bf x}_{0}(s)-{\bf x}_{0}(s-s'))-\omega_{\bf k}s'
\right)}\langle e^{i{\bf k}\cdot{\bf\Delta\bf x}(s,s-s')}\rangle 
\vspace{3mm} \\
\ds  \qquad + \frac{1}{2}\frac{d}{dt}\frac{e^{2}}{m_{e}^{2}}
\int_{0}^{t}ds\int_{0}^{s}ds'\sum_{\bf k}|\delta E_{\parallel\bf k}|^{2}
e^{i\left({\bf k}\cdot({\bf x}_{0}(s)-{\bf x}_{0}(s-s'))-\omega_{\bf k}s'
\right)}\langle e^{i{\bf k}\cdot{\bf\Delta\bf x}(s,s-s')}\rangle.
\ea\!\!\!
\label{eq:r54} 
\ee

Note that dif\/ferent expressions ought to be used for ${\bf\Delta\bf x}
(s,s-s')$ for each of the last two integrals of expression (\ref{eq:r54}).
Then using equation (\ref{eq:r16}) and def\/ining the following dif\/fusion
coef\/f\/icient we obtain 
\[
\ba{l}
\ds \tilde{\bf{\cal D}}_{\bf\perp\bf\perp}      =  \tilde{\bf
D}_{\bf\perp\bf\perp}+\frac{{\bf D}_{v_{\perp}\perp}}{\Omega_{e}},
 \vspace{3mm}\\
\ds \underline{\tilde{\cal D}}_{\parallel\perp}  = 
\underline{D}_{v_{\parallel}\perp}+\frac{\underline{D}_{
v_{\parallel}v_{\perp}}}{\Omega_{e}}.
\ea
\label{eq:r54z} 
\]

\noindent
{\bf Case when $s'>0$}
\be
\ba{l}
\ds \exp{\left(-\frac{1}{2}\langle\left[{\bf
k}\cdot{\bf\Delta\bf x}(s,s-s')\right]^{2}\rangle\right)}
 =
\exp{\left\{
-\frac{k_{\parallel}^{2}}{3}D_{v_{\parallel}v_{\parallel}}{s'}^2(3s-2s')
\right. }
\vspace{3mm}\\
\ds \qquad {\left. -k_{\parallel}\left[{s'}^2\underline{\tilde{\cal
D}}_{\parallel\perp}(s-s')+s^{2}s'{\partial\underline{\tilde{\cal
D}}_{\parallel\perp}(s-s')\over\partial s}\right]\cdot{\bf k}_{\perp}
\right\}} 
\vspace{3mm}\\
\ds \qquad \times\exp{\left\{-{\bf
k}_{\perp}\cdot\left[\Omega_{e}^{2}{s'}^2(s-s')\tilde{\bf
D}_{\bf\perp\bf\perp}(s-s')+ s'\left(s-\frac{s'}{2}\right)\frac{d}{ds}\tilde{\bf
D}_{\bf\perp\bf\perp}(s-s')\right.\right.}
\vspace{3mm} \\
\ds \qquad  {\left.\left.+{s'}^2{\partial\tilde{\bf {\cal
D}}_{\bf\perp\bf\perp}(s-s')\over\partial s}\right] \cdot{\bf
k}_{\perp}\right\}} .
\ea
\label{eq:r55} 
\ee

\noindent
{\bf Case when $s'<0$}
\be
\ba{l}
\ds \exp{\left(-\frac{1}{2}\langle\left[{\bf
k}\cdot{\bf\Delta\bf x}(s,s-s')\right]^{2}\rangle\right)}
 =\exp{\left\{
-\frac{k_{\parallel}^{2}}{3}D_{v_{\parallel}v_{\parallel}}{s'}^2(3s-s')
\right.}
\vspace{3mm}\\
\ds \qquad {\left. -
k_{\parallel}\left[{s'}^2\underline{\tilde{\cal
D}}_{\parallel\perp}(s)-s'(s-s')^{2}{\partial\underline{\tilde{\cal
D}}_{\parallel\perp}(s)\over\partial s}\right]\cdot{\bf k}_{\perp}\right\}}
\vspace{3mm} \\
\ds \qquad  \times\exp{\left\{-{\bf
k}_{\perp}\cdot\left[\Omega_{e}^{2}{s'}^2s\tilde{\bf
D}_{\bf\perp\bf\perp}(s)- s'\left(s-\frac{s'}{2}\right)\frac{d}{ds}\tilde{\bf
D}_{\bf\perp\bf\perp}(s)\right. \right.}
\vspace{3mm}\\
\ds \qquad {\left. \left. +{s'}^2{\partial\tilde{\bf {\cal
D}}_{\bf\perp\bf\perp}(s)\over\partial s}\right] \cdot{\bf
k}_{\perp}\right\}} .
\ea
\label{eq:r56} 
\ee
Using the last two expressions (\ref{eq:r55}) and (\ref{eq:r56}), and
changing variables in the second integral, we obtain the following
expression for the parallel dif\/fusion coef\/f\/icient 
\[
\ba{l}
\ds D_{v_{\parallel}v_{\parallel}}=\frac{e^{2}}{m_{e}^{2}}\sum_{\bf
k}|\delta E_{\parallel\bf k}|^{2}\int_{0}^{t}ds\; \cos{\left({\bf k}\cdot({\bf
x}_{0}(t)-{\bf x}_{0}(t-s))-\omega_{\bf k}s\right)} 
\vspace{3mm} \\
\ds \qquad \times  \exp{\left\{
-\frac{k_{\parallel}^{2}}{3}D_{v_{\parallel}v_{\parallel}}s^{2}(3t-2s)-
k_{\parallel}\left[s^{2}\underline{\tilde{\cal
D}}_{\parallel\perp}(t-s)+t^{2}s{\partial\underline{\tilde{\cal
D}}_{\parallel\perp}(t-s)\over\partial t}\right]\cdot{\bf k}_{\perp}
\right\}} 
\vspace{3mm} \\
\ds \qquad  \times\exp{\left\{-{\bf
k}_{\perp}\cdot\left[\Omega_{e}^{2}s^{2}(t-s)\tilde{\bf
D}_{\bf\perp\bf\perp}(t-s)+ s(t-\frac{s}{2})\frac{d}{dt}\tilde{\bf
D}_{\bf\perp\bf\perp}(t-s)\right.\right.}
\vspace{3mm} \\
\ds \qquad  {\left.\left.+s^{2}{\partial\tilde{\bf{\cal
D}}_{\bf\perp\bf\perp}(t-s)\over\partial t}\right]\cdot{\bf
k}_{\perp}\right\}}
\ea
\label{eq:r57}
\]
A similar calculation for the perpendicular components of the
dif\/fusion tensor as well as the cross components, leads to
\[
\ba{l}
\ds \tilde{\bf D}_{\bf\perp\bf\perp}(t)  =  \frac{c^{2}}{B^{2}}\sum_{\bf
k}({\bf\delta\bf E}_{\bf k}\times\hat{\bf z})({\bf\delta\bf
E}_{\bf k}^{*}\times\hat{\bf z})\int_{0}^{t}ds\; \cos{\left({\bf k}\cdot({\bf
x}_{0}(t)-{\bf x}_{0}(t-s))-\omega_{\bf k}s\right)} 
\vspace{3mm} \\
\ds \qquad \times  \exp{\left\{
-\frac{k_{\parallel}^{2}}{3}D_{v_{\parallel}v_{\parallel}}s^{2}(3t-2s)-
k_{\parallel}\left[s^{2}\underline{\tilde{\cal
D}}_{\parallel\perp}(t-s)+t^{2}s{\partial\underline{\tilde{\cal
D}}_{\parallel\perp}(t-s)\over\partial t}\right]\cdot{\bf k}_{\perp}
\right\}} 
\vspace{3mm} \\                               
\ds \qquad \times  \exp{\left\{-{\bf
k}_{\perp}\cdot\left[\Omega_{e}^{2}s^{2}(t-s)\tilde{\bf
D}_{\bf\perp\bf\perp}(t-s)+ s(t-\frac{s}{2})\frac{d}{dt}\tilde{\bf
D}_{\bf\perp\bf\perp}(t-s)\right.\right.}
\vspace{3mm} \\
\ds \qquad  + {\left.\left.s^{2}{\partial\tilde{\bf{\cal
D}}_{\bf\perp\bf\perp}(t-s)\over\partial t}\right]\cdot{\bf
k}_{\perp}\right\}} 
\ea
\label{eq:r58}
\]
and similar expressions for the cross dif\/fusion coef\/f\/icients.

It is clear from these results that the resonance broadening ef\/fects due to
scattering of electrons by the Modif\/ied Two Stream Farley-Buneman
waves can not be accounted for by just replacing $\omega$ by
$\omega+ik_{\perp}^{2}D^{*}$ in the resonant part of the dispersion
relation. It is obvious from the expressions above that there is a
complex time dependence of the dif\/fusion coef\/f\/icients. Moreover, most
of the published work (Dum and Dupree~[7], Sudan~[19], 
Robinson~[16], Robinson and Honary~[17]) ignores the cross correlation
and consequently the cross dif\/fusion.

\setcounter{equation}{0}

\section{The Nonlinear Dielectric Function} 
\label{sec-epsilon}
In order to obtain the dielectric function one has to use the results of the
previous section. The expression for the f\/luctuating part of the electron
distribution function (\ref{eq:r13}) can now be written in the following form
after using equation (\ref{eq:r53})

\newpage

\be
\ba{l}
\ds \delta f_{e\bf k\omega}({\bf v})  =  -i{\cal L}_{1{\bf k}\omega}\langle
f_{e}\rangle  \vspace{3mm} \\
\ds \qquad \times \int_{t_{0}}^{t}dt' \; e^{\left(i\omega
(t-t')-i{\bf k}\cdot({\bf x}_{0}(t)-{\bf x}_{0}(t'))\right)}\exp{\left\{-
\frac{k_{\parallel}^{2}}{3}D_{v_{\parallel}v_{\parallel}}(t^{3}+{t'}^2(2t'
-3t))\right\}}  
  \vspace{3mm} \\
\ds \qquad \times \exp{\left\{-{\bf
k}_{\perp}\cdot\left(\frac{(t^{2}-{t'}^2)}{2}\frac{d}{dt'}\tilde{\bf
D}_{\bf\perp\bf\perp}(t')+(t-t')^{2}{\partial\tilde{\bf{\cal
D}}_{\bf\perp\bf\perp}(t')\over \partial t'}\right.\right.} 
 \vspace{3mm}\\
\ds \qquad + \left.\left.\Omega_{e}^{2} (t-t')^{2}t'\tilde{\bf
D}_{\bf\perp\bf\perp}(t')\right)\cdot{\bf k}_{\perp}-
k_{\parallel}\left(\underline{\tilde{\cal
D}}_{\parallel\perp}(t')(t-t')^{2} \right.\right. 
\vspace{3mm} \\
\ds \qquad  + {\left.\left.t^{2}(t-t'){\partial\underline{\tilde{\cal
D}}_{\parallel\perp}(t')\over\partial t'}\right)\cdot{\bf
k}_{\perp}\right\}}
\ea
\label{eq:n1} 
\ee
using this expression (\ref{eq:n1}) along with equation (\ref{eq:i4}) in
Poisson's equation, and changing the integration variable from $t'$ to
$s=t-t'$ leads to the nonlinear dielectric function 
\be
\ba{l}
\ds \epsilon({\bf k},\omega) =
1-\frac{\omega_{pi}^{2}}{\omega^{2}-k^{2}\frac{
T_{i}}{m_{i}}+i\nu_{in}\omega}-i\frac{\omega_{pe}^{2}}{k^{2}}\int
d{\bf v} \left({\bf k}\cdot{\bf
v}_{\perp}\frac{1}{v_{\perp}}{\partial\over\partial
v_{\perp}}+k_{\parallel}{\partial\over\partial v_{\parallel}}\right. 
\vspace{3mm}\\
\ds \qquad  +  \left. \frac{{\bf
k}\times\hat{\bf z}}{\Omega_{e}}\cdot\tilde{\bf\nabla}_{\perp}\right)\langle
f_{e}\rangle\int_{0}^{t}ds \; e^{\left(i\omega
s-i{\bf k}\cdot({\bf x}_{0}(t)-{\bf x}_{0}(t-s))\right)} 
\exp{\left\{-\frac{k_{\parallel}^{2}}{3}
D_{v_{\parallel}v_{\parallel}}s^{2}(3t-2s) \right\}}
\vspace{3mm} \\
\ds \qquad \times 
\exp{\left\{-
k_{\parallel}\left(s^{2}\underline{\tilde{\cal
D}}_{\parallel\bf\perp}(t-s)+t^{2}s{\partial\underline{\tilde{\cal
D}}_{\parallel\perp}(t-s)\over\partial t}\right)\cdot {\bf k}_{\perp}
\right\}}
\vspace{3mm} \\
\ds \qquad \times \exp{\left\{-{\bf
k}_{\perp}\cdot\left(s(t-\frac{s}{2}
)\frac{d}{dt}\tilde{\bf D}_{\bf\perp\bf\perp}(t-s)+\Omega_{e}^{2}
s^{2}(t-s)\tilde{\bf D}_{\bf\perp\bf\perp}(t-s)\right.\right.} 
\vspace{3mm} \\
\ds \qquad  + {\left.\left.s^{2}{\partial\tilde{\bf{\cal
D}}_{\bf\perp\bf\perp}(t-s)\over\partial t}\right)\cdot{\bf
k}_{\perp}\right\}} .
\ea \hspace{-15.53pt}
\label{eq:n2} 
\ee

This dispersion relation dif\/fers considerably from that used by
Robinson and Honary [17] and Robinson~[16] in many aspects, and
therefore the consequences on the physics of irregularities in the
auroral as well as equatorial E regions are signif\/icantly
dif\/ferent. We will show in the next section how the accurate
resonance broadening calculation af\/fects the threshold for
Farley-Buneman instablity. We will also show how it af\/fects the
important problem of aspect angles. 

In order to extract the
information hidden in the dispersion relation we have 
to evaluate the time integral in the expression (\ref{eq:n2}), a nontrivial
calculation.

A f\/inal note on the time dependence. As mentioned earlier in the
paper, the time dependence that appears in the right hand side of the
dispersion relation is a slow time dependence necessary for energy
and momentum balance. In other words, the time dependence is
necessary for wave saturation. This is a classical problem. Linear
theory ignores the slow time dependence and predicts a time
independent growth rate which in turn suggests that waves will grow
indef\/inetly. Quasi-linear theory remedies this critical problem by
introducing a slow time dependence in the background distribution 
function. We have retained both the fast and the slow time dependence
and Fourier transformed over the fast time scale. The slow time scale
is associate with the dif\/fusion time scale in the classical
quasi-linear theory. This same time scale reappears in the resonance
broadening analysis.

\setcounter{equation}{0}

\section{The Farley-Buneman Case}
\label{sec-FBthresholds}

In order to simplify the results and obtain a direct comparison with the
classical Farley-Buneman results we will assume that $\delta E_{\parallel}=0$,
and the we have isotropic turbulence. This allows us to eliminate the parallel
and cross dif\/fusion ef\/fects, and use a simplif\/ied dispersion relation
\[
\ba{l}
\ds \epsilon({\bf k},\omega) =
1-\frac{\omega_{pi}^{2}}{\omega^{2}-k^{2}\frac{
T_{i}}{m_{i}}+i\nu_{in}\omega}-i\frac{\omega_{pe}^{2}}{k^{2}}\int
d{\bf v} \left({\bf k}\cdot{\bf
v}_{\perp}\frac{1}{v_{\perp}}{\partial\over\partial
v_{\perp}}+\frac{{\bf k}\times\hat{\bf
z}}{\Omega_{e}}\cdot\tilde{\bf\nabla}_{\perp}\right)\langle f_{e}\rangle 
\vspace{3mm} \\
\ds \qquad \times \int_{0}^{t}ds\;
e^{\left(i\omega s-i{\bf k}\cdot({\bf x}_{0}(t)-{\bf x}_{0}(t-s))\right)}
\exp{\left\{-k_{\perp}^{2}\left(s(t-\frac{s}{2} )\frac{d}{dt}
D^{*}(t-s)\right.\right.}
\vspace{3mm}\\
\ds \qquad  +  {\left.\left.\Omega_{e}^{2}
s^{2}(t-s)D^{*}(t-s)+s^{2}{\partial\tilde{\cal
D}_{\bf\perp\bf\perp}(t-s)\over\partial t}\right)\right\}},
\ea
\label{eq:FB1}
\]
where we have assumed an isotropic spectrum for simplicity, and
replaced ${\bf D}_{\perp\perp}$ by $D^{*}$. Then assuming a slow time
dependence of the dif\/fusion coef\/f\/icient $D^{*}$ we can further
simplify the expression for the dielectric function to obtain
\[
\ba{l}
\ds \epsilon({\bf k},\omega) =
1-\frac{\omega_{pi}^{2}}{\omega^{2}-k^{2}\frac{
T_{i}}{m_{i}}+i\nu_{in}\omega}-i\frac{\omega_{pe}^{2}}{k^{2}}\int
d{\bf v} \left({\bf k}\cdot{\bf
v}_{\perp}\frac{1}{v_{\perp}}{\partial\over\partial
v_{\perp}}+\frac{{\bf k}\times\hat{\bf
z}}{\Omega_{e}}\cdot{\bf\nabla}_{\perp}\right)\langle f_{e}\rangle
\vspace{3mm}\\
\ds \qquad \times  \int_{0}^{t}ds \; e^{\left(i\omega s-i{\bf k}\cdot({\bf
x}_{0}(t)-{\bf x}_{0}(t-s))\right)}\exp{\left\{-k_{\perp}^{2}\Omega_{e}^{2}
s^{2}(t-s)D^{*}(t-s)\right\}}.
\ea
\label{eq:FB2}
\]
At this point one can explicitly express the unperturbet orbits of the
electrons in terms of Bessel functions, that is
\[
\ba{l}
\ds e^{\left(i\omega s-i{\bf k}\cdot({\bf x}_{0}(t)-{\bf
x}_{0}(t-s))\right)}  =  
\sum_{n,m=-\infty}^{+\infty}J_{n}\left(k_{\perp}\rho_{\perp}\right)
J_{m}\left(k_{\perp}\rho_{\perp}\right) 
\vspace{3mm}\\
\ds \qquad \times 
\exp{\left\{i\left(\omega-{\bf k}\cdot{\bf v}_{E}^{(0)}-n\Omega_{e}\right)s+
i(n-m)\phi-i(n-m)\Omega_{e} t\right\}} 
\ea
\label{eq:FB3}  
\]
the integral over the velocity allows us to reduce the double sum to a single
sum by integrating over $\phi$ to obtain
\[
\ba{l}
\ds \epsilon({\bf k},\omega) =
1-\frac{\omega_{pi}^{2}}{\omega^{2}-k^{2}\frac{
T_{i}}{m_{i}}+i\nu_{in}\omega}-i\frac{\omega_{pe}^{2}}{k^{2}}\int
v_{\perp}\; dv_{\perp}\sum_{n=-\infty}^{+\infty}J_{n}^{2}(k_{\perp}\rho_{\perp})
\vspace{3mm} \\
\ds \qquad \times R(\omega-{\bf k}\cdot{\bf
v}_{E}^{(0)}-n\Omega_{e})\left(\frac{n\Omega_{e}}{v_{\perp}}
{\partial\over\partial v_{\perp}}+\frac{{\bf k}\times\hat{\bf
z}}{\Omega_{e}}\cdot\tilde{\bf\nabla}_{\perp}\right)\langle
f_{e}\rangle, 
\ea
\label{eq:FB4}
\]
where the resonance function $R$ is given by
\begin{equation}
\ba{l}
\ds R(\omega-{\bf k}\cdot{\bf
v}_{E}^{(0)}-n\Omega_{e})
\vspace{3mm}\\
\ds \qquad =\int_{0}^{t}ds \; \exp{\left\{i\left(\omega-{\bf
k}\cdot{\bf v}_{E}^{(0)}-n\Omega_{e}\right)s -k_{\perp}^{2}\Omega_{e}^{2}
s^{2}(t-s)D^{*}(t-s)\right\}}.
\ea
\label{eq:FB5}
\end{equation}

It is clear from the expression (\ref{eq:FB5}) that the resonance
function is similar to that of the unmagnetized case derived by
Ishihara {\it et al.}~[12] and 
Salat~[18]. Therefore one can extract similar properties of the time
integral and consequently obtain some useful information regarding the
broadening of the wave-particle resonance. If we were to make the same change
of variables
\[
\ds \tau_{K} = (k_{\perp}^{2}\Omega_{e}^{2}D^{*})^{-\frac{1}{3}},
\qquad \ds T =  \frac{t}{\tau_{K}}, \qquad
\ds U = \omega-{\bf k}\cdot{\bf v}_{E}^{(0)}-n\Omega_{e}, \qquad
\ds \tilde{R} =  \frac{R}{\tau_{K}}
\label{eq:FB7}
\]
then the resonance function can be written as
\[
\tilde{R}(U,T)=\int_{0}^{T} dS\; \exp{\left\{iUS-S^{2}(T-S)\right\}}.
\label{eq:FB8}
\]
Note that only the real part of $R$ appears in the expressions for the
dif\/fusion tensor components.

The f\/inal question to be addressed regarding this problem is that
related to collisions between electrons and neutrals. It has been
established through linear theory, Kadomtsev~[13], Coppi and
Rosenbluth~[2] and Hendel {\it et al.}~[9], that collisional
ef\/fects enter the dispersion relation through the introduction of a
collisional damping $d_{e}^{coll}$ depending on the wave frequency,
and which can be expressed as follows for electron neutral collisions
\[
d_{e}^{coll}=\frac{k^{2}v_{e}^{2}}{2(\nu_{en}-i\omega)}=k_{\perp}^{2}
D_{e}^{coll},  \label{eq:FB6} 
\]
where $D_{e}^{coll}$ is the collisional dif\/fusion
coef\/f\/icient.

When combining the last two equations, we obtain the correct
dispersion relation for Farley-Buneman waves. To be more specif\/ic we can
integrate the Vlasov equation with the Bhatnagar collision operator,
one can formally solve for the distribution function and therefore the charge
dewnsity to obtain the dispersion relation. The steps of this procedure are
described below. 

We start with the collisional Vlasov equation, which we write in the
following form
\[
\frac{d\delta f_{e}}{dt}+\nu_{en}\delta
f_{e}=-\frac{e}{m_{e}}{\bf\nabla}\Phi\cdot{\bf\nabla}_{\bf v}f_{e0}+\nu_{en}
\frac{f_{0}}{n_{0}}\delta n({\bf x},t),
\label{eq:a1}
\]
where the total time derivative $d/dt$ is the derivative along the perturbed
particle orbits. The solution to this equation can be written in an integral
form
\[
\ba{l}
\ds \delta f_{e}  =  f_{e0}({\bf v})e^{-\nu_{en}t}
\vspace{3mm}  \\
\ds \qquad  \!\!+  \nu_{en}e^{-\nu_{en}t}\!
\int_{0}^{t}\!dt' \left\{-\frac{e}{m_{e}\nu_{en}}{\bf\nabla} \Phi({\bf
x}(t'),t')\cdot{\bf\nabla}_{\bf v}f_{e0}({\bf v})+ \frac{f_{0}({\bf
v})}{n_{0}}\delta n({\bf x}(t'),t')\right\} e^{-\nu_{en}t'}
\ea \!
\label{eq:a2} 
\]
writing
\[
\ds \delta f_{e}({\bf x},{\bf v},t) = \delta f_{e{\bf
k}\omega}\exp\{i\left[{\bf k}\cdot{\bf x}(t)-\omega t\right]\},
  \qquad
\ds \Phi({\bf x}, t) =  \Phi_{{\bf k}\omega}\exp\{i\left[{\bf
k}\cdot{\bf x}(t)-\omega t\right]\}
\label{eq:a3}
\]
leads to
\be
\ba{l}
\ds \delta f_{e{\bf k}\omega}  =  f_{e0}\exp\{-i\left[{\bf k}\cdot{\bf
x}(t)-(\omega+i\nu_{en})t\right]\} 
\vspace{3mm} \\
\ds \qquad +  \int_{0}^{t}dt^{'}\left\{i\frac{e
\Phi_{{\bf k}\omega}}{T_{e}}f_{e0}{\bf k}\cdot{\bf v}+\frac{f_{0}}{n_{0}}
\delta n_{{\bf k}\omega}\right\}e^{i(\omega+i\nu_{en})(t-t^{'})}\langle
e^{-i\left[{\bf k}\cdot({\bf x}(t)-{\bf x}(t^{'}))\right]}\rangle,
\ea
\label{eq:a4} 
\ee
where ${\bf x}(t)={\bf x}_{0}+{\bf\delta x}(t)$, with ${\bf x}_{0}$
representing the unperturbed orbits, while ${\bf\delta x}$ is the
perturbation due to the random electric f\/ields. Note that we have
assumed that $f_{e0}$ is a drifting Maxwellian. Neglecting the f\/irst term in
equation (\ref{eq:a4}) for long times $t\rightarrow\infty$, and using the
results of the previous section we obtain
\[
\ba{l}
\ds \delta f_{e{\bf k}\omega}  =  \int_{0}^{t}ds\left\{i\frac{e
\Phi_{{\bf k}\omega}}{T_{e}}f_{e0}{\bf k}\cdot{\bf
v}+\nu_{en}\frac{f_{0}}{n_{0}} \delta n_{{\bf k}\omega}\right\} 
\vspace{3mm} \\
\ds \qquad \times e^{i\left((\omega+i\nu_{en})(t-t')-{\bf
k}\cdot({\bf x}_{0}(t)-{\bf
x}_{0}(t-s))\right)}\exp{\left(-\frac{1}{2}\langle\left[{\bf
k}\cdot{\bf \Delta x}(t,t-s)\right]^{2}\rangle\right)}.
\ea
\label{eq:a5} 
\]
Using the results of the previous section where the Resonance function was
calculated, we can deduce the expression for the f\/luctuating part of the
distribution function, and then integrate over velocity to obtain an
expression for the density
\[
\ba{l}
\ds \frac{{\bf\delta n}_{\bf k\omega}}{n_{0}} = i\frac{e\Phi_{{\bf
k}\omega}} {T_{e}}\int d{\bf v}\int_{0}^{t}ds\;\frac{f_{e0}}{n_{0}}{\bf
k}\cdot{\bf v} e^{i\left((\omega+i\nu_{en})(t-t')-{\bf k}\cdot({\bf
x}_{0}(t)-{\bf x}_{0}(t-s))\right)} 
\vspace{3mm} \\
\ds \qquad \times \exp{\left(-\frac{1}{2}\langle\left[{\bf
k}\cdot{\bf \Delta x}(t,t-s)\right]^{2}\rangle\right)} 
\vspace{3mm}\\
\ds \qquad  +  \nu_{en}\frac{{\bf\delta n}_{\bf 
k\omega}}{n_{0}}\int d{\bf v}\int_{0}^{t}ds\; \frac{f_{0}}{n_{0}}
e^{i\left((\omega+i\nu_{en})(t-t')-{\bf k}\cdot({\bf x}_{0}(t)-{\bf
x}_{0}(t-s))\right)} \vspace{3mm} \\
\ds \qquad \times 
\exp{\left(-\frac{1}{2}\langle\left[{\bf k}\cdot{\bf \Delta
x}(t,t-s)\right]^{2}\rangle\right)}.
\ea
\label{eq:a6} 
\]
Substituting this expression along with the expression for the ion charge
density in Poisson's equation we obtain the dispersion relation
\[
\ba{l}
\ds \epsilon({\bf k},\omega)  = 
1-\frac{\omega_{pi}^{2}}{\omega^{2}-k^{2}\frac{T_{i}}{m_{i}}+i\nu_{in}\omega}
\vspace{3mm}\\ 
\ds \qquad  \!\!\!\!+  \frac{i}{k^{2}\lambda_{De}^{2}}\frac{
\int d{\bf v}\int_{0}^{t}ds\frac{f_{e0}}{n_{0}}{\bf k}\cdot{\bf v}
e^{i\left((\omega+i\nu_{en})(t-t')-{\bf k}\cdot({\bf x}_{0}(t)-{\bf
x}_{0}(t-s))-\frac{1}{2}\langle\left[{\bf k}\cdot{\bf \Delta
x}(t,t-s)\right]^{2}\rangle\right)}}{\left(1-\nu_{en}\int d{\bf
v}\int_{0}^{t}ds \frac{f_{0}}{n_{0}}
e^{i\left((\omega+i\nu_{en})(t-t')-{\bf k}\cdot({\bf x}_{0}(t)-{\bf
x}_{0}(t-s))-\frac{1}{2}\langle\left[{\bf k}\cdot{\bf \Delta
x}(t,t-s)\right]^{2}\rangle\right)}\right)}.
\ea
\label{eq:eq:a7}
\]

\section{Summary}

We have shown that the components of the electron dif\/fusion
coef\/f\/icient are time dependent and the conventional result that
suggested replacing $\omega$ by $\omega+ik_{\perp}^{2}D^{*}$ in the
resonant part of dielectric function is not valid. This alters
a number of results obtained through the application of the classical
resonance broadening calculations of Dum and Dupree~[7], such as
Sudan's results~[19] 
and Robinson's results~[17] concerning the thresholds of the
Farley-Buneman and Gradient drift instabilities in the ionosphere.
The correct results will be presented in a subsequent paper to be
submitted in the near future. We have aslo added the parallel
dif\/fusion as well as the cross dif\/fusion coef\/f\/icients. 
The former is identical to the one dimensional analog derived by
Ishihara {\it et al.}~[12] and Salat~[18], the latter has never been
calculated explicitly. 

Finally, we have explicitely derived the dispersion relation for the
Farley-Buneman waves using the improved resonance broadening
formalism. Further details on the Farley-Buneman thresholds and
transport will be published in the near future.

\subsection*{Acknowledgement}

The author is grateful to J.-P. St-Maurice and D.R. Moorcroft for a
number of discussions related to Farley-Buneman turbulence in the
ionosphere, and would also like to thank A.~Hirose for his deep comments.

Funding for this research has been provided by NSERC, a Canadian Research 
funding agency.

\label{hamza-lp}


\begin{thebibliography}{99}

\footnotesize

\bibitem{B.1}  Buneman O., {\it Phys. Res. Lett.}, 1963, V.10, 285.

\bibitem{C.1} Coppi B. and Rosenbluth M.N., Plasma Physics and Controlled
Nuclear Fusion Research, International Atomic Energy Agency, Vienna,
1966, Vol.1, p.628.

\bibitem{C.2} Cook I. and  Sanderson A.D., {\it Plasma Phys.},
1974, V.16, 977.

\bibitem{D.1} Drummond W.E. and Pines D., {\it Nucl. Fusion Suppl.}, 
1962, V.2, 1049.

\bibitem{D.2} Dupree T.H., {\it Phys. Fluids}, 1966, V.9.

\bibitem{D.3} Dupree T.H., {\it Phys. Fluids}, 1968, V.11.

\bibitem{D.4}  Dum C.T. and  Dupree T.H., {\it Phys. Fluids}, 1970,
V.13.

\bibitem{F.1} Farley D.T., {\it J. Geophys. Res.}, 1963, V.68, 6083.

\bibitem{H.1} Hendel H.W., Coppi B., Perkins F. and Politzer P., {\it
Phys. Rev. Lett.}, 1967, V.18, 439.

\bibitem{I.1} Ishihara O. and Hirose A., {\it Phys. Fluids}, 1985,
V.28, 2159.

\bibitem{I.2} Ishihara O., Grabowski C. and Hirose A., {\it Phys.
Fluids B}, 1990, V.2, 270.

\bibitem{I.3} Ishihara O., Xia X. and Hirose A., {\it Phys. Fluids
B},  1992, V.4, 349.

\bibitem{K.1} Kadomtsev B.B.,  Plasma Turbulence, Ch.4, Academic Press Inc.,
New York, 1965.

\bibitem{K.2} Kleva R.B., {\it Phys. Fluids B}, 1991, V.3, 3312.

\bibitem{R.1} Rolland P., {\it J. Plasma Phys.}, 1976, V.15, 57.

\bibitem{R.2} Robinson T.R., {\it J. Atmos. Terr. Phys.}, 1986, V.48, 417.

\bibitem{R.3} Robinson T.R. and Honary F., {\it J. Geophys. Res.},
1990, V.95, 1073. 

\bibitem{S.1} Salat A., {\it Phys. Fluids}, V.31, 1499.

\bibitem{S.2} Sudan R.N., {\it J. Geophys. Res.}, 1983, V.88, 4853.

\bibitem{W.1} Weinstock J., {\it Phys. Fluids}, 1969, V.12.

\bibitem{Y.1} Kleva R.B. and Drake J.F., {\it Phys. Fluids}, 1986,
V.27, 1686. 

\end{thebibliography}
\end{document}